# Flat-band driven bipolar anomalous Nernst effect towards volt-scale transverse thermoelectric conversion

Haofu Wen[1]*, Xiaodong Guo[2]*, Xiaokang Xu[1]*, Peng Di[1]*, Xuebiao Ji[1], Ruoqi Wang[1], Shijun Yuan[1], Kunquan Hong[1], Qionghua Zhou[1]✉, Qian Chen[1]✉, Taishi Chen[1]✉, Jinlan Wang[1]✉

1. *Key Laboratory of Quantum Materials and Devices of Ministry of Education, School of Physics, Southeast University, Nanjing 211189, China*
2. *Wuhan National High Magnetic Field Center School of Physics, Huazhong University of Science and Technology, Wuhan 430074, China*
3. *Suzhou Laboratory, Suzhou 215004, China*

** Haofu Wen, Guo Xiaodong[2]*, Xiaokang Xu, and Peng Di contributed equally.*

✉ *Corresponding author. E-mail: chentaishi@seu.edu.cn, qc119@seu.edu.cn, qh.zhou@seu.edu.cn, jlwang@seu.edu.cn*

## Abstract

Unipolar anomalous Nernst thermopiles are critically plagued by metallic interconnects, which induce heavy Joule dissipation and compromise the module's packing density, ultimately capping voltage generation. Here, we circumvent these limitations by establishing an intrinsic bipolar anomalous Nernst effect (BANE) paradigm, enabling seamless series integration analogous to Seebeck $n$–$p$ pairs without passive metallic bridges. Focusing on flat-band topological systems, through machine learning assisted screening across multiple target properties, we identified 249 BANE materials, exhibiting near-room-temperature polarity switching. Guided by this predictive framework, we experimentally realized this phenomenon in single crystals of the Weyl ferromagnet, $Co_2TiGe$. Strikingly, while its intrinsic anomalous Hall conductivity (AHC) remains rigorously temperature-invariant up to the Curie temperature, the anomalous Nernst signal undergoes a sharp polarity inversion across $T \approx 225$ K. First-principles calculations attribute this behavior to the interplay between flat bands and Weyl nodes, where thermal depopulation shifts the chemical potential across the band edge, inverting the sign of $\partial\sigma_{xy}/\partial\varepsilon$ via the Mott relation. This work opens an avenue toward compact, monolithic transverse thermopiles capable of volt-scale generation free from interfacial parasitic dissipation.

## Introduction

Thermopiles operating on the anomalous Nernst effect (ANE) present a transformative architecture for waste-heat harvesting[1-3]. By virtue of an orthogonal cross-plane geometry, the ANE fundamentally decouples the driving thermal gradient from the generated electrical field[4-7]. This planar configuration eliminates the fragile three-dimensional leg assemblies intrinsic to conventional Seebeck modules, permitting the conformal deposition of sub-micrometre ferromagnetic films directly onto non-planar, curved thermal interfaces[7,8]. Driven by advances in topological band theory, magnetic nodal semimetals have dramatically escalated the room-temperature Nernst thermopower ($S_{xy}$) beyond $6\ \mu\mathrm{V\ K^{-1}}$, pushing output potentials into the millivolt regime[9-14]. To accelerate the discovery of such quantum materials, high-throughput screening approaches, ranging from first-principles calculations to machine learning, have been employed[9,15]. For example, Sakai et al. used high-throughput calculations to guide the discovery of a giant ANE in earth-abundant, non-toxic $D0_3$-phase $Fe_3Ga$[9].

Nevertheless, solely pursuing larger anomalous Nernst coefficients represents an inherently limited trajectory that cannot bridge the gap to practical, volt-scale power generation, leaving transverse thermopiles lagging far behind their commercially mature Seebeck counterparts. This conventional pursuit is critically hindered by two fundamental bottlenecks. First, at the microscopic level, the Mott relation dictates that the anomalous Nernst conductivity is governed by the energy derivative of the intrinsic anomalous Hall conductivity (AHC) at the Fermi energy, $\alpha_{xy} \propto -(\pi^2 k_\mathrm{B}^2 T/3e)\big(\partial\sigma_{xy}/\partial\varepsilon\big)\,|_{\varepsilon_\mathrm{F}}$[16,17]. While theoretically sign-reversible, realizing such a polarity inversion in conventional dispersive or Weyl bands requires shifting the Fermi level across broad AHC profiles ($\Delta E_F > 100\ \mathrm{meV}$)[18]. Experimentally, this has historically mandated aggressive chemical substitution or heavy atomic doping[19]. Crucially, such extrinsic chemical disorder severely broadens electronic states, introduces detrimental defect scattering, and often destroys the delicate topological band crossings, rendering non-destructive ANE polarity control an insurmountable challenge. Furthermore, the intrinsic AHC in known magnetic topological semimetals has rapidly converged toward an empirical ceiling of $\sim 10^3$ S/cm, offering rapidly diminishing returns for further single-band magnitude engineering[9,10,12,20-23]. Second, and most debilitating, as a direct consequence of this doping dilemma, existing transverse thermopiles are universally restricted to unipolar materials[7,9-13,23,24]. Because adjacent active elements generate transverse voltages with identical polarity, cascading them in series strictly necessitates an equivalent array of passive metallic interconnects to route the current loop. These parasitic conductive tracks introduce prohibitive interfacial contact resistances, induce severe Joule dissipation, and consume vital module footprint, drastically reducing the active packing density and capping the volumetric power output[3,25].

Here, we circumvent these long-standing barriers by conceiving and demonstrating a monolithic transverse thermopile architecture based on the intrinsic BANE. We show that flat bands in time-reversal-symmetry-broken ferromagnets concentrate Berry curvature into an exceptionally narrow energy window, producing an abrupt change in the energy-resolved AHC spectrum[9,10,12,26,27]. According to the Mott relation, shifting the chemical potential across the summit of this localized AHC peak directly reverses the sign of $\partial\sigma_{xy}/\partial\varepsilon$, thereby triggering a natural polarity inversion of the transverse thermoelectric response, a condition that can be precisely targeted via delicate band-filling control, such as gentle plasma-assisted ion implantation. To explore the generality of this mechanism, we combined machine-learning-assisted predictions with a hierarchical screening strategy based on key properties, identifying 249 promising candidates for near-room-temperature BANE. As a definitive experimental realization, we pulled high-purity, centimeter-scale single crystals of the Heusler Weyl ferromagnet $Co_2TiGe$ via the Czochralski method[28]. Strikingly, whereas its moderate intrinsic AHC ($\sim 40$ S cm$^{-1}$) remains virtually temperature-independent up to the Curie point ($T_C \approx 400$ K), the anomalous Nernst signal undergoes a sharp polarity reversal at $T \approx 225$ K. First-principles calculations reveal that this transition originates from thermal depopulation shifting the chemical potential across a sharp AHC extremum produced by Weyl nodes embedded within nearly dispersionless flat bands. By enabling seamless, bridge-free serial cascading within a single continuous material platform, this work charts an actionable path toward monolithic, ultra-compact transverse thermopiles free of interfacial parasitic dissipation.

**Results**

Fig. 1 introduces the concept of volt-scale BANE thermopiles and the machine-learning-assisted screening for materials discovery. To transcend the geometric and resistive bottlenecks that inherently constrain conventional transverse thermoelectric conversion devices, we conceive a monolithic, flexible thermopile architecture enabled by BANE, shown in Fig.1a. In this design, Nernst building blocks with opposing response polarities, designated as $n$- and $p$-type legs, are arranged sequentially on an electrically insulating yet thermally conductive flexible substrate. These alternating thermoelectric domains are joined directly in a seamless, head-to-tail junction, entirely bypassing passive metallic interconnects. This monolithic configuration enables uninterrupted serial voltage accumulation, directly replicating the polarity-alternating circuit topology of classical Seebeck unicouples while eradicating dead volumes, interfacial contact resistances, and parasitic thermal mismatches[1,2]. Moreover, this planar transverse paradigm readily facilitates double-sided conformal integration across both faces of an ultrathin flexible membrane, quadrupling the cumulative open-circuit voltage per unit projected footprint and delivering an order-of-magnitude leap in volumetric power density relative to conventional unipolar counterparts.

Realizing such a bipolar transverse response necessitates an unconventional electronic structure wherein the intrinsic AHC exhibits a colossal, sign-reversing energy derivative within an exceptionally narrow chemical potential window. We formalize this criterion via a flat-band engineering strategy (Fig. 1b–d). When two weakly dispersive bands with bandwidths ($W$) comparable to the room-temperature thermal energy ($W \approx k_{\mathrm{B}}T_{\mathrm{room}} \approx 26$ meV) approach the Fermi level, three canonical electronic topologies can be rationally engineered (Fig. 1b): topologically non-trivial crossings generated by band inversion (left), accidental degeneracies in topologically trivial regimes (middle), or avoided crossings opened by spin–orbit coupling (SOC) that enforce ultranarrow gaps (right). A unifying feature of these configurations is the singular localization of momentum-space Berry curvature ($\Omega_z(\boldsymbol{k})$), which produces a pronounced AHC peak sharply confined within a thermal window of $\sim 26$ meV (Fig. 1c). In the finite-temperature transport regime governed by the generalized Mott relation, translating the chemical potential across this sharp AHC inverts the sign of the energy derivative, $\partial\sigma_{xy}/\partial\varepsilon$. Consequently, this slope reversal dictates opposing polarities for the transverse thermoelectric conductivity $\alpha_{xy}$, establishing distinct $n$-type ($\alpha_{xy} < 0$) and $p$-type ($\alpha_{xy} > 0$) anomalous Nernst operational regimes within an identical, single-phase crystallographic host (Fig. 1d).

Guided by these design criteria, a search is conducted for magnetic materials that host topological flat bands near the Fermi level while retaining itinerant carriers for electrical transport. Starting from an initial pool of 32,006 flat-band materials, high-throughput screening is performed based on topological, metallic, and magnetic properties. In the first screening step, based on topology and metallicity, the pool is narrowed to 3,836 intermediate candidates. Notably, magnetic labels are missing for approximately half of the materials in the candidate pool. Therefore, an XGBoost classifier is trained to distinguish magnetic from nonmagnetic materials, further reducing the pool to 1,211 candidates. Finally, 249 promising BANE candidates are identified, as shown in Fig. 1e and Extended Data Fig.1.

Among the identified candidates, the full-Heusler ferromagnet $Co_2TiGe$ stands out as a quintessential platform to experimentally demonstrate this flat-band BANE mechanism, shown in Extended Data Fig. 1. Full-Heusler compounds provide an ideal materials playground characterized by high Curie temperatures and extensive chemical flexibility for isoelectronic tuning[29]. Crucially, symmetry analysis indicates that $Co_2TiGe$ hosts magnetic Weyl nodal networks that generate gigantic, energy-concentrated Berry curvature hot spots[28,30]. However, realizing macroscopic, defect-free single crystals of titanium-bearing Heusler systems has historically remained a formidable metallurgical hurdle due to incongruent melting behavior and severe peritectic elemental segregation. Overcoming these synthetical bottlenecks, we successfully pulled high-purity, centimetre-scale $Co_2TiGe$ single crystals using an ultra-high-temperature tetra-arc Czochralski technique (Extended Data Fig. 2).

Fig. 2a illustrates the anomalous Hall effect (AHE) and BANE response in single crystal bulk $Co_2TiGe$. To unambiguously correlate transverse charge and thermoelectric transport, we concurrently measured the AHE and ANE on high-purity $Co_2TiGe$ single crystals within an orthogonal experimental geometry (Fig. 2a). The longitudinal thermal gradient or electric current was oriented along the $[100]$ axis ($-\nabla_x T \parallel j_x \parallel [100]$), an external magnetic field was applied along the out-of-plane easy axis ($B \parallel [001]$), and the corresponding transverse responses were detected simultaneously along the $[010]$ axis ($E_y \parallel [010]$). Field-dependent Hall resistivity measurements ($\rho_{xy}(B)$) across a broad temperature window reveal canonical ferromagnetic anomalous Hall characteristics (Fig. 2b). Throughout the long-range ordered ferromagnetic state below the Curie temperature ($T_{\mathrm{C}} \approx 400$ K), the spontaneous anomalous Hall resistivity $\rho_{xy}^{\mathrm{A}}$ maintains a strictly unipolar character, retaining a single sign while growing monotonically with temperature until it collapses precipitously near $T_{\mathrm{C}}$.

Strikingly, the transverse thermoelectric transport departs dramatically from this monotonic, sign-invariant anomalous Hall behavior. Isothermal anomalous Nernst thermopower curves ($S_{xy}(B)$) unveil a pronounced BANE traversing the cryogenic to elevated-temperature regimes (Fig. 2c). Upon warming from $10$ K, $S_{xy}$ initially rises to a distinct positive local maximum of $\sim +0.10\,\mu\mathrm{V\,K^{-1}}$ around $180\,\mathrm{K}$, subsequently subsides toward zero near $220$ K, and abruptly reverses its polarity into a rapidly escalating negative signal. Isofield scans collected under a saturating magnetic field of $\mathrm{B} = 2\ \mathrm{T}$ track this continuous thermal trajectory, definitively pinning the sign-reversal temperature at $T \approx 225$ K (Fig. 2d).

This distinct qualitative dichotomy, wherein the AHE remains rigorously sign-invariant while the anomalous Nernst thermopower undergoes a dramatic polarity inversion within the same crystallographic phase, provides decisive physical insight. Because extrinsic skew-scattering or side-jump mechanisms typically couple the transverse channels directly to the net magnetization and longitudinal resistivities, they are fundamentally incapable of inducing an isolated sign reversal in $S_{xy}$ without a concomitant sign flip in $\rho_{xy}$. Instead, this decoupling constitutes the smoking-gun experimental signature of the intrinsic Berry curvature mechanism[20,21]. In the framework of the Mott relation, whereas the AHC integrates the Berry curvature over all occupied states below the Fermi level, the anomalous Nernst conductivity $\alpha_{xy}$ probes its sharp energy derivative $(\partial\sigma_{xy}/\partial\varepsilon)$ at the chemical potential. Consequently, this remarkable polarity inversion provides an unyielding empirical benchmark for accurately tracking the thermal drift of the actual Fermi level and mapping the underlying nodal band dispersion in $Co_2TiGe$.

Fig. 3 shows pinpointing the Fermi level in $Co_2TiGe$ via dual transport constraints. We first extracted the field-dependent transverse Hall conductivity $\sigma_{xy}(B)$ across diverse temperatures (Fig. 3a) via the standard inversion of the resistivity tensor, $\sigma_{xy} = -\frac{\rho_{xy}}{\rho_{xx}^2+\rho_{xy}^2}$. In stark contrast to the pronounced thermal variation in the Hall resistivity, the spontaneous AHC $\sigma_{xy}^{\mathrm{A}}$ remains remarkably temperature-invariant across the entire ferromagnetic state below $T_c$. This robustness fundamentally manifests the high crystallographic perfection of our single crystals, where extrinsic skew-scattering and side-jump contributions are largely suppressed. Consequently, the transverse charge transport is overwhelmingly governed by the intrinsic Berry curvature integrated over the occupied Bloch sea. Linearly extrapolating the high-field Hall conductivity to zero field ($B \to 0$) yields a plateaued spontaneous AHC value of $\sigma_{xy}^{\mathrm{A}} \approx +40\ \mathrm{S\ cm^{-1}}$across the entire thermal span (Fig. 3b). Projecting this experimental value onto the zero-temperature AHC spectrum $\sigma_{xy}(\varepsilon)$ computed via the Berry-curvature Kubo formula (Fig. 1c) yields two degenerate candidate Fermi energies within the energy window of $+0.2$ to $+0.5$ eV: $\varepsilon_1 \approx 0.2480$ eV and $\varepsilon_2 \approx 0.4473$ eV relative to the nominal density functional theory (DFT) Fermi level (Fig. 3b, inset). Crucially, the static scalar magnitude of $\sigma_{xy}$ alone cannot lift this energetic degeneracy.

To resolve this ambiguity, we turned to the anomalous Nernst conductivity $\alpha_{xy}$, which sensitively monitors the energy derivative of AHC at the Fermi level. We rigorously converted the thermopower data into the transverse thermoelectric conductivity tensor (Fig. 3c) through the complete relation: $\alpha_{xy} = \frac{\rho_{xx}S_{xy}-\rho_{xy}S_{xx}}{\rho_{xx}^2+\rho_{xy}^2} \approx \frac{S_{xy}}{\rho_{xx}} - \frac{\rho_{xy}S_{xx}}{\rho_{xx}^2}$, where $S_{xy}$, $\rho_{xx}$, $S_{xx}$, and $\rho_{xy}$ were synchronously measured on the identical single-crystal platform (seen in Extend Data Fig. 3). The sign-reversal trajectory persists unabated after this full tensor transformation, definitively ruling out spurious artifacts from longitudinal thermopower cross-coupling ($S_{xx}$) or carrier-mobility variations. The saturated transverse thermoelectric conductivity reaches an amplitude of $\alpha_{xy} \approx 0.3$ A m$^{-1}$ K$^{-1}$. To establish an absolute quantitative benchmark against finite-temperature DFT simulations that account for Fermi–Dirac thermal broadening, Fig. 3d tracks the spontaneous, normalized anomalous Nernst conductivity $\alpha_{xy}/T$ as a function of temperature. Following the polarity switch at $T \approx 210$ K, $\alpha_{xy}/T$ undergoes a steep, monotonic escalation, peaking at $\sim -0.8 \times 10^{-3}$ A m$^{-1}$ K$^{-2}$ near 350 K prior to the thermal collapse of long-range ferromagnetism at $T_c$. Theoretical evaluations of $\alpha_{xy}/T$ at the two candidate energies reveal that while $\varepsilon_1 = 0.2480$ eV yields a monotonic, sign-invariant profile, the simulated thermal trajectory at $\varepsilon_2 = 0.4473$ eV captures both the precise compensation temperature 210 K and the absolute quantitative magnitude of the high-temperature peak (Fig. 3d, inset). By simultaneously reconciling the sign-invariant $\sigma_{xy}^{A} \approx +40$ S cm$^{-1}$ with the delicate temperature-dependent sign-reversal profile of $\alpha_{xy}/T$, we unequivocally pin the actual Fermi level of our as-grown $Co_2TiGe$ at $E_F = 0.4473$ eV. The temperature evolution of the finite-temperature anomalous Nernst conductivity tensor is presented in Extended Data Figs. 5 and 6.

Fig. 4 shows first-principles DFT calculations substantiate the microscopic quantum origin of the observed BANE in $Co_2TiGe$, also shown in Extended data Fig.4. Exactly at the experimentally determined Fermi level ($\varepsilon - E_F^0 = +0.4473$ eV), the calculated electronic structure unveils a topologically non-trivial crossing embedded within weakly dispersive flat bands characterized by an ultranarrow bandwidth ($W \approx 10$ meV) (Fig. 4a). High-resolution scanning across the full Brillouin zone confirms that this flat-band manifold hosts 8 symmetry-related pairs of magnetic Weyl nodes possessing opposite chiralities. As a direct consequence of the bulk-boundary correspondence, these nodes project distinct, open topological Fermi arcs onto the semi-infinite $(010)$ cleavage surface (Fig. 4b). Wilson-loop spectra tracking the evolution of Wannier charge centres across spherical manifolds enclosing individual nodes yield quantized winding numbers, rigorously verifying their topological chiral charges of $C = \pm 1$ (Fig. 4c).

Momentum-resolved Berry curvature calculations ($\Omega_z(k)$) along the flat-band crossing delineate a colossal divergence exceeding -$10^5$ $Å^2$ in the immediate vicinity of the Weyl singularities (Fig. 4e). Crucially, this singular Berry curvature collapses precipitously within an ultraconfined energy window of merely $\pm 5$ meV around the crossing apex. This sharp energy localization microscopically generates the quasi-$\delta$-function AHC resonance at $+0.4473$ eV and fundamentally dictates the onset of BANE. Concurrently, a broad, dispersive Berry curvature background on the order of $\sim 10^4$ $Å^2$ extends along high-symmetry paths (notably along ***Γ–L***). Although an order of magnitude smaller in local amplitude than that of the flat-band Weyl hotspots, its expansive phase-space integration over the entire Brillouin zone, show in Extended Data Fig. 4 and Supp. Fig. 3.

In stark contrast to the AHE[20,21], the anomalous Nernst conductivity is governed by the energy derivative of the Hall conductivity at the chemical potential rather than its absolute integrated magnitude[16,17]. The ultrafast quench of the -$10^5$ $Å^2$ Berry curvature within the sub-10 meV flat-band regime generates an exceptional derivative anomaly ($|\partial\sigma_{xy}/\partial\varepsilon| \gg 0$). Consequently, as finite-temperature thermal broadening or minute chemical potential shifts traverse this singular resonance, the abrupt sign inversion of $\partial\sigma_{xy}/\partial\varepsilon$ directly unleashes the colossal BANE response observed in $Co_2TiGe$.

**Summary and Discussion**

In conclusion, by exploiting the energy derivative singularity at flat-band edges, this work establishes a transformative paradigm that fundamentally overcomes the long-standing "doping dilemma" in Mott-driven transverse thermoelectrics, decisively elevating output capabilities from the sub-millivolt deadlock into the volt-scale regime. Conventional approaches in dispersive and Weyl semimetals are inherently trapped: their broad AHC profiles dictate that reversing the anomalous Nernst polarity requires heavy chemical substitution ($\Delta E_F > 100$ meV)[18], which inevitably introduces severe lattice disorder, degrades carrier mobility, and annihilates topological band features. In stark contrast, we have shown that the divergent DOS in flat-band ferromagnets condenses Berry curvature into an exceptionally narrow energy window. The resulting ultra-steep slope ($\partial\sigma_{xy}/\partial\varepsilon$) enables clean, intrinsic polarity reversal under minute thermodynamic shifts ($\Delta E_F < 26$ meV) without extrinsic chemical doping.

Taking the prototypical flat-band ferromagnet $Co_2TiGe$ DOS as an experimental benchmark, it delivers a peak $n$-type transverse thermoelectric coefficient $S_{xy} \approx -0.7\ \mu\mathrm{V\ K^{-1}}$ at 350 K, a low electrical resistivity of $\rho \approx 0.2\ \mathrm{m\Omega \cdot cm}$, and robust Hall and Seebeck responses (Extended Data Note 7 and Extended Data Table I). By matching it with its symmetric $p$-type counterpart via subtle chemical tuning, monolithic series cascading, analogous to conventional longitudinal Seebeck $p$–$n$ couples, entirely eliminates inactive metallic interconnects and parasitic interfacial contact resistance. Consequently, a $1\ \mathrm{cm} \times 1\ \mathrm{cm}$ flexible BANE module yields a record-high open-circuit voltage exceeding 4.0 V under a modest out-of-plane temperature gradient (Extended Data Note 7 and Extended Data Fig. 7). Crucially, backed by machine-learning screening that identified 249 room-temperature candidate materials, our strategy is neither an isolated material anomaly nor a modest parameter tuning. Instead, it unlocks the unexploited derivative dimension of flat bands, bridging microscale topological singular physics directly to bridge-free, high-voltage monolithic thermoelectric generation.

## Methods

### Single-Crystal Growth

Single crystals of $Co_2TiGe$ were grown by the Czochralski pulling method in an arc-discharge furnace. Stoichiometric amounts of high-purity Co, Ti, and Ge pieces (molar ratio $2:1:1$) were first loaded into a mono-arc furnace. The chamber was evacuated to a high vacuum better than $1 \times 10^{-5}$ Pa and baked for 10 hours for thorough outgassing, followed by backfilling with high-purity argon (99.98%) to ambient pressure. Before melting the constituents, a pure titanium getter was arc-melted several times to scavenge residual traces of oxygen. The $Co_2TiGe$ ingot was flipped and remelted six times to ensure compositional homogeneity, yielding a dense polycrystalline button. The resulting polycrystalline button was subsequently transferred into a tetra-arc Czochralski crystal growth furnace (Quantum Design). The growth chamber was purged five times with ultrahigh-purity argon, evacuated to a base pressure below $1 \times 10^{-7}$ Pa, and held under high vacuum for $12$ h. After backfilling with high-purity argon to $1$ atm, the single-crystal boule was continuously pulled over a duration of 3 hours, producing an ingot approximately $2$ cm in length and $4.5$ mm in diameter, shown in the inset of Extended data Fig. 2a.

### Sample Preparation and Characterization

To reduce structural defects and improve atomic site ordering, the as-grown $Co_2TiGe$ crystals were sealed in evacuated quartz ampoules backfilled with high-purity argon, annealed at $800^{\circ}C$ for two weeks, and naturally furnace-cooled. The crystallographic orientations were determined via back-reflection X-ray Laue photography (Photonic Science, UK). Oriented single-crystal specimens with well-defined prismatic geometries were sliced along specific crystallographic axes using a high-precision, custom-built wire electrical discharge machine. Chemical composition and spatial uniformity were verified by scanning electron microscopy equipped with energy-dispersive X-ray spectroscopy (SEM-EDS) shown in Extended Data Fig. 3.

**Magnetic and Thermoelectric Measurements**

The Seebeck and Nernst coefficients were measured under high vacuum in a Physical Property Measurement System (PPMS, Quantum Design) using a chip-resistor heater and two Chromel--constantan (type ~ E) thermocouples for differential temperature measurements. The heating current was supplied by a current source (Keithley ~ 6221), and the thermoelectric voltage was measured using a DC nanovoltmeter (Keithley ~ 2182A). At each fixed magnetic field, data were acquired using a three-step background--signal--background sequence, with the heater switched off, on, and off, respectively, to reduce the effects of background voltage and its drift.

**Machine-learning classification model**

The magnetic/nonmagnetic classifier was implemented with the XGBoost Python package[31]. The dataset comprised 65,838 magnetic materials and 62,558 nonmagnetic materials from Materials Project v2026.04.13[32]. Each sample was represented by a 481-dimensional feature vector constructed using Magpie[33] and CHGNet[34]. The dataset was randomly divided into training, validation, and test sets at a 70:10:20 ratio. Model selection used the validation set, and performance was evaluated on the held-out test set using ROC-AUC, accuracy, and F1 score at a classification threshold of 0.5.

**First-Principles Calculations**

The geometry optimization and electronic structure calculations in this work were carried out with projector-augmented wave (PAW)[35] methods within Vienna ab initio Simulation Package (VASP)[36,37]. The exchange-correlation potential was adopted by the generalized gradient approximation (GGA) of Perdew-Burke-Ernzerhof (PBE) functional[38-42]. The cutoff energy was set as 500 eV. The energy and force convergence criteria were set to be $10^{-9}$ eV and $10^{-4}$ eV/Å, respectively. For geometry optimization, the first Brillouin zone was sampled by Γ-centered k mesh of 12×12×12. The optimized lattice constant is 5.840 Å. Correlation effects for the Co-*3d* electrons were treated by the DFT +U ($U_{eff}$ = 0.3 eV) method. SOC was included self-consistently. Ti Co $s$, $p$, $d$ orbitals and Ge $s$, $p$ orbitals were used to generate a tight-binding Hamiltonian constructed by Wannier90 package[43] with the maximally localized Wannier functions (MLWF)[44]. The momentum-space Berry curvature was calculated on a dense 401×401×1 k-point mesh. The energy-dependent intrinsic AHC ($\sigma_{xy}$) were evaluated by dense k-point 71×71×71 integration across the Brillouin zone.

## Figures and Captions

**Fig. 1 |The concept of volt-scale BANE thermopiles and the machine-learning-assisted screening for materials discovery. a,** Schematic illustration of the monolithic BANE device fabricated on a flexible polyimide substrate. Alternating $p$- and $n$-type transverse thermoelectric segments are directly and seamlessly concatenated in-plane without extrinsic metallic interconnects, enabling constructive voltage superposition and macroscopic power accumulation under a uniform out-of-plane thermal gradient. **b,** Three representative topological band archetypes engineered to generate narrow-bandwidth flat bands: a nodal flat band (left), an accidental-crossing flat band (middle), and an avoided-crossing flat band with an ultranarrow bandgap (right). **c,** Energy-resolved AHC and density of states profiles produced by the three flat-band topologies, both exhibiting singular, rapid variations across the Fermi level ($E_{\mathrm{F}}$). Driven by the Mott relation ($\alpha_{xy} \propto -\partial\sigma_{xy}/\partial\varepsilon$), these sharp spectral variations serve as definitive physical descriptors for identifying large transverse thermoelectric responses. The full width at half maximum of the AHC peak (FWHM $= 26$ meV) closely matches the thermal energy scale at room temperature ($k_{\mathrm{B}}T \approx 25.8$ meV), establishing a vital quantitative criterion for realizing robust BANE sign switching at finite temperatures. **d,** Temperature-dependent shift of the chemical potential traversing the sharp AHC resonance, which drives an abrupt polarity reversal in the anomalous Nernst conductivity tensor ($\alpha_{xy}$) and underpins the microscopic origin of the finite-temperature BANE. **e,** Machine learning-assisted screening funnel for identifying 249 candidate materials from the Initial candidate pool.

**Fig. 2 | Intrinsic AHE and BANE in $Co_2TiGe$ single crystals. a,** Schematic illustration of the transverse transport measurement geometry. An in-plane temperature gradient $-\nabla T \parallel [100]$ and an out-of-plane magnetic field $B \parallel [001]$ generate a mutually orthogonal transverse electric field $E_y \parallel [010]$. The anomalous Hall resistivity $\rho_{xy}$ is measured using the identical transverse lead geometry, ensuring consistent spatial registration across both galvanomagnetic and thermoelectric channels.

**b,** Magnetic field dependence of the Hall resistivity $\rho_{xy}(B)$ recorded at representative temperatures up to $400$ K, displaying robust spontaneous anomalous Hall response. **c,** Magnetic field dependence of the anomalous Nernst thermopower $S_{xy}(B)$ across selected temperatures, showcasing sharp hysteretic switching and an unambiguous sign inversion between $220$ K and $250$ K. **d,** Temperature evolution of the saturated anomalous Nernst thermopower $S_{xy}(T)$ under $\mu_0 H = 2$ T. A clear zero-crossing polarity reversal occurs at 225 K, corroborating the BANE mechanism driven by the thermal displacement of the chemical potential across the sharp Berry curvature resonance. At temperatures above $350$ K, $|S_{xy}|$ precipitously decays toward zero due to thermal demagnetization and the subsequent collapse of long-range ferromagnetic order approaching the Curie temperature ($T_C \approx 400$ K).

**Fig. 3 | Quantitative determination of the actual Fermi level via experimental transport and DFT calculations. a,** Magnetic field dependence of the Hall conductivity $\sigma_{xy}(B)$ at representative temperatures, derived from the tensor inversion $\sigma_{xy} = -\rho_{xy}/(\rho_{xx}^2 + \rho_{xy}^2)$. Below the Curie temperature ($T_C$), the saturated Hall conductivity exhibits an exceptionally weak temperature dependence and stabilizes at approximately $40$ S cm$^{-1}$, pointing to an intrinsically driven topological origin. **b,** Temperature dependence of the AHC $\sigma_{xy}^A$, extracted via linear extrapolation of high-field data to zero field ($B \to 0$). Inset: DFT-calculated zero-temperature AHC profile $\sigma_{xy}(\varepsilon)$ within an energy window of $0.2$ to $0.5$ eV, featuring sharp twin resonance peaks with an intrinsic maximum of $40$ S cm$^{-1}$, in striking agreement with the low-temperature experimental plateau. **c,** Experimentally determined anomalous Nernst conductivity $\alpha_{xy}(B)$ at various temperatures, converted via the full thermoelectric tensor relation $\alpha_{xy} = S_{xy}/\rho_{xx} - \rho_{xy}S_{xx}/\rho_{xx}^2$. **d,** Temperature evolution of the normalized anomalous Nernst conductivity $\alpha_{xy}/T$ recorded under $\mu_0 H = 2$ T. Inset: DFT-simulated finite-temperature $\alpha_{xy}/T$ as a function of temperature across the identical thermal range. The exquisite quantitative and line-shape agreement unambiguously pinpoints the actual Fermi energy at a calibrated rigid shift of $\Delta E_F =$

$+0.4473$ eV relative to the pristine DFT level.

**Fig. 4 | Weyl topology and energy-sensitive Berry curvature in $Co_2TiGe$ a,** Electronic band structure calculated along the off-symmetry momentum paths $P_1$–$P_2$ and $P_3$–$P_4$, purposefully selected to traverse the pronounced Berry curvature hot spots generated by spin–orbit coupling (SOC)-induced band crossings. **b,** Surface-projected band structure along the $\overline{P}_1$–$\overline{\Gamma}$–$\overline{P}_2$ path on the semi-infinite (010) surface, displaying topologically non-trivial Fermi arcs connecting the surface projections of opposite-chirality Weyl nodes. Red and blue circles denote the projected Weyl points $W^+$ and $W^-$ with positive and negative chiral charges, respectively. **c,** Wilson-loop spectra tracking the evolution of Wannier charge centres calculated on closed spherical manifolds enclosing the individual SOC-induced crossing points. The opposite integer winding directions rigorously verify chiral topological charges of $C = +1$ and $-1$ for the $W^+$ and $W^-$ Weyl nodes, respectively (inset). **d–f,** Momentum-resolved Berry curvature ($\Omega_z$) distributions evaluated along the representative path in a across three closely spaced chemical potentials: $E_\mathrm{F} = 0.4423$ eV(d), $0.4473$ eV(e), and $0.4523$ eV(f), corresponding to incremental energy shifts of -5 meV, 0, and +5 meV relative to the target level. Driven by the topological flat-band dispersion, the Berry curvature undergoes singular spectral reconstructions within an ultranarrow energy window of merely a few millielectronvolts. Inset: The aligned positions of $E_\mathrm{F}$ superimposed onto the sharp resonance peak of the AHC profile $\sigma_{xy}(\varepsilon)$, demonstrating the microscopic origin of the steep $\partial\sigma_{xy}/\partial\varepsilon$ slope that underpins the robust BANE polarity reversal.

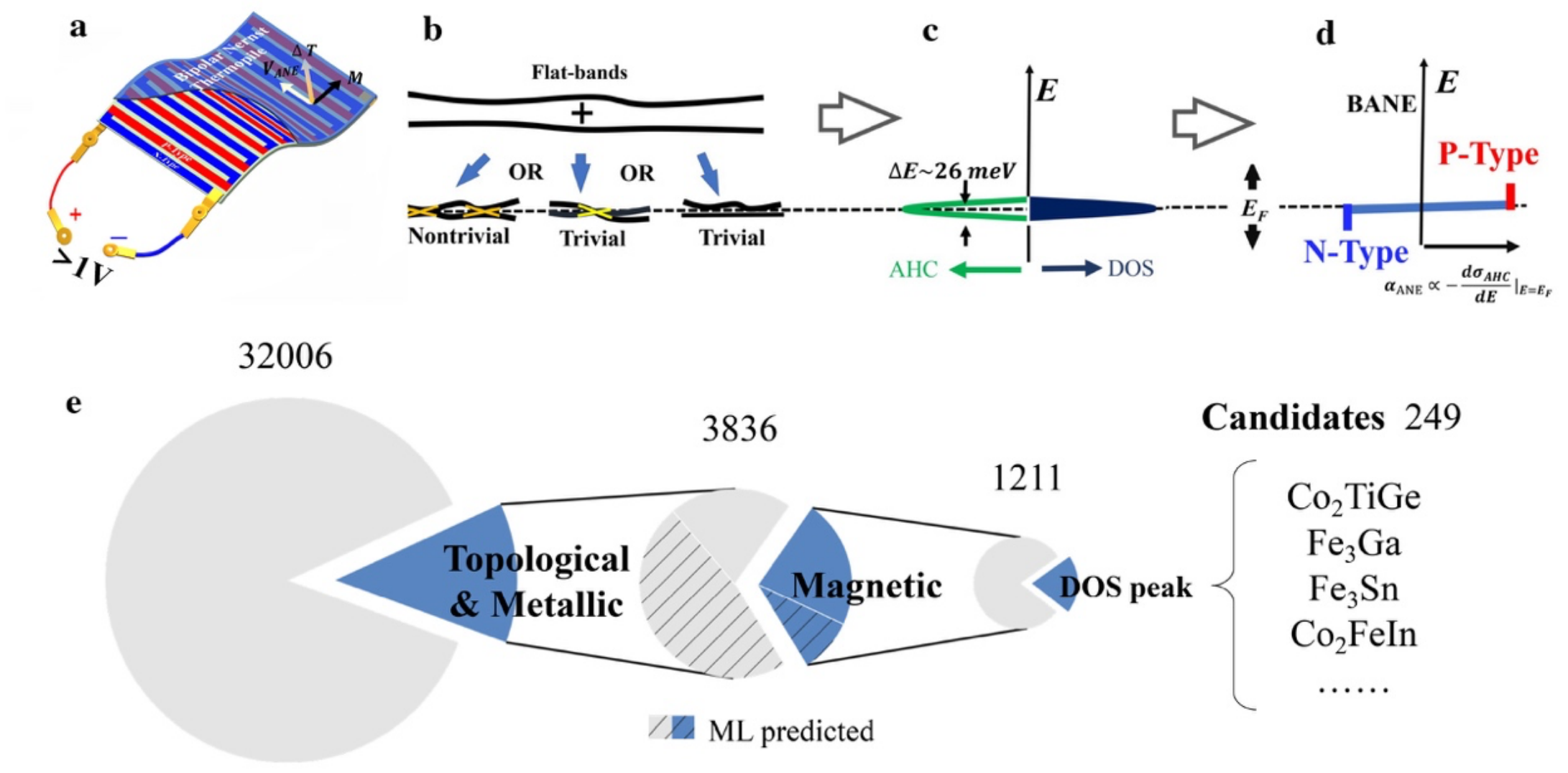

a
b
c
d
Flat-bands
OR
OR
Nontrivial
Trivial
Trivial
E
ΔE~26 meV
AHC
DOS
$E_F$
BANE
E
P-Type
N-Type
32006
e
3836
1211
Candidates 249
Topological & Metallic
Magnetic
DOS peak
$Co_2TiGe$
$Fe_3Ga$
$Fe_3Sn$
$Co_2FeIn$
……
ML predicted

Fig. 1

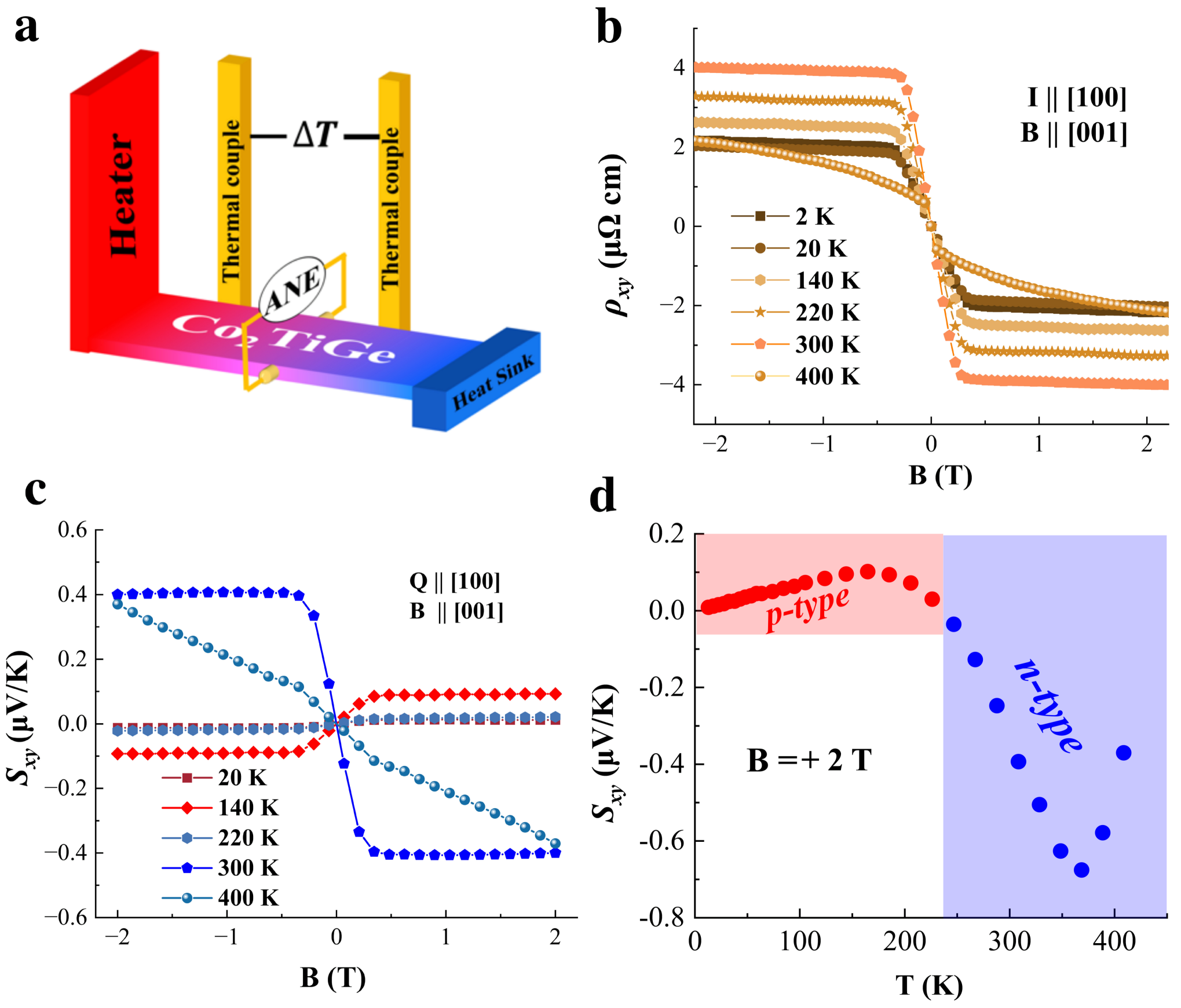

a
Heater
Thermal couple
ΔT
Thermal couple
ANE
Co2TiGe
Heat Sink
b
I || [100]
B || [001]
$\rho_{xy}$ (μΩ cm)
2 K
20 K
140 K
220 K
300 K
400 K
B (T)
c
Q || [100]
B || [001]
$S_{xy}$ (μV/K)
20 K
140 K
220 K
300 K
400 K
B (T)
d
p-type
n-type
B = + 2 T
$S_{xy}$ (μV/K)
T (K)

Fig. 2

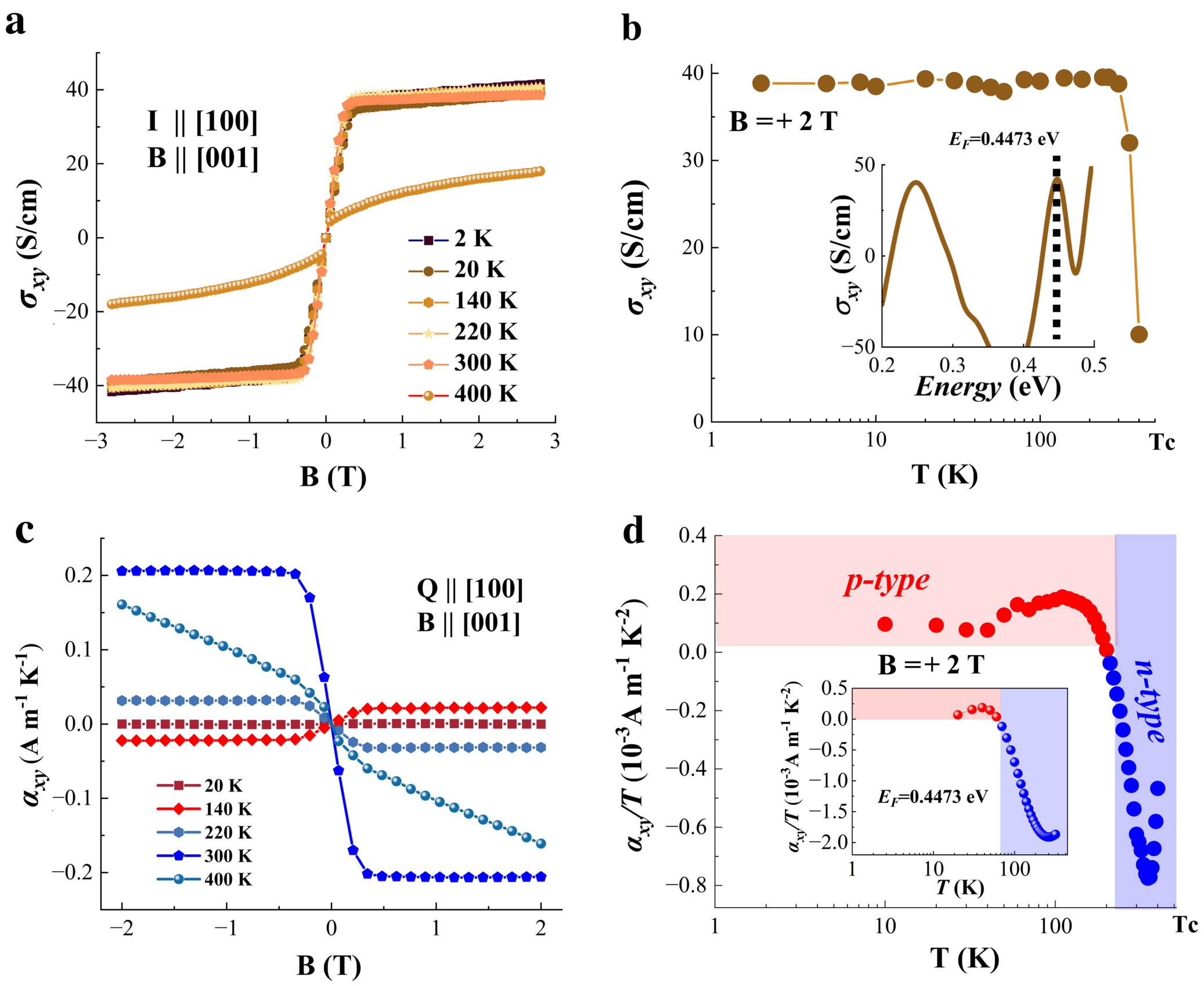
a
I || [100]
B || [001]
2 K
20 K
140 K
220 K
300 K
400 K
B (T)
b
B =+ 2 T
$E_F$=0.4473 eV
Energy (eV)
T (K)
Tc
c
Q || [100]
B || [001]
20 K
140 K
220 K
300 K
400 K
B (T)
d
p-type
n-type
B =+ 2 T
$E_F$=0.4473 eV
T (K)
Tc

Fig. 3

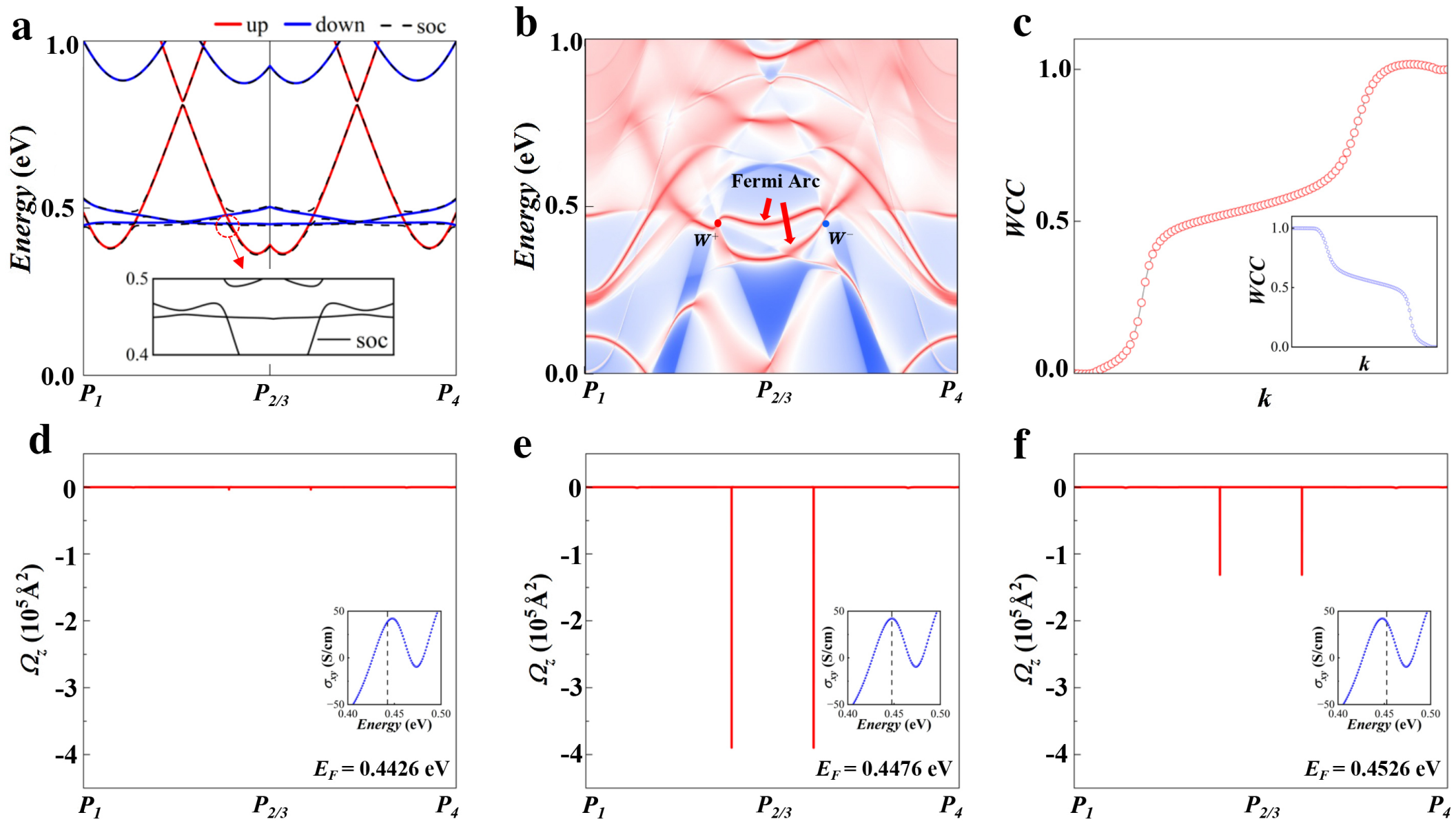
a
up
down
soc
Energy (eV)
$P_1$
$P_{2/3}$
$P_4$
b
Fermi Arc
W
c
WCC
k
d
$E_F$ = 0.4426 eV
e
$E_F$ = 0.4476 eV
f
$E_F$ = 0.4526 eV

Fig. 4

**Acknowledgements** We thank Prof. Song Bao, Dr. Zihang Song, and Prof. Jinsheng Wen from Nanjing University for their generous assistance with the single-crystal Laue diffraction and orientation measurements. This work was supported by the National Key Research and Development Program of China (Grant No. 2023YFA1406600 to T.C.) and the National Natural Science Foundation of China (Grant No. 12274068 to T.C., T2321002 to J.W., 22373013 to Q.Z.), the National Natural Science Foundation of China (Grant 22573013). The authors are thankful for the computational resources from the Big Data Computing Center of Southeast University. The authors thank the computational resources from the Big Data Computing Center of Southeast University, and the Center for Fundamental and Interdisciplinary Sciences of Southeast University. The authors thank the Center for Fundamental and Interdisciplinary Sciences of Southeast University for the support in magnetization measurements.

**Author contributions** H.W., X.X., P.D. and X.G. contributed equally to this work. T.C., Q.C., Q.Z. and J.W. conceived and supervised the project. T.C., H.W. and X.G. planned the experiments. J.W., Q.Z. and P.D. performed the machine-learning computational screening. T.C. and H.W. grew the single crystals and prepared the samples. X.G., T.C. and H.W. carried out the transport and magnetization measurements and analyzed the data. Q.C. and X.X. performed the first-principles calculations. H.W. performed the chemical analyses. T.C. wrote the manuscript with input from all authors. All authors discussed the results and commented on the manuscript.

**Competing interests** The authors declare no competing interests.

## Extended Data

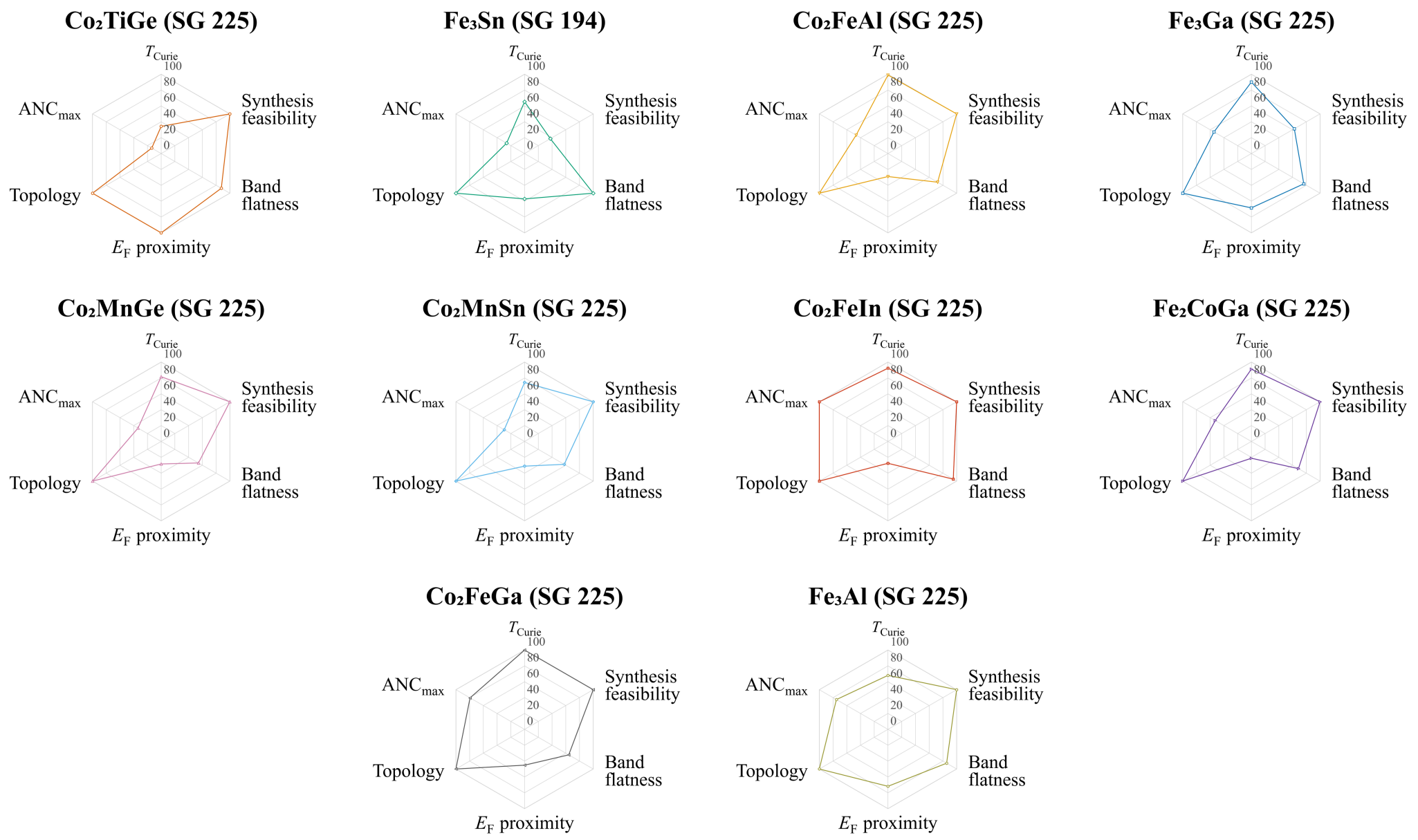

**Extended Data Fig. 1 |**Multi-criteria comparison of ten anomalous Nernst candidate materials. Radar plots show scores (0–100) for Curie temperature, synthesis feasibility informed by reported syntheses and experimental experience, band flatness, flat-band proximity to the Fermi level and topological screening based on the Materials Flatband Database[26] and maximum absolute anomalous Nernst conductivity (ANC) from calculations and experiments. Higher scores indicate more favourable performance for the corresponding criterion.

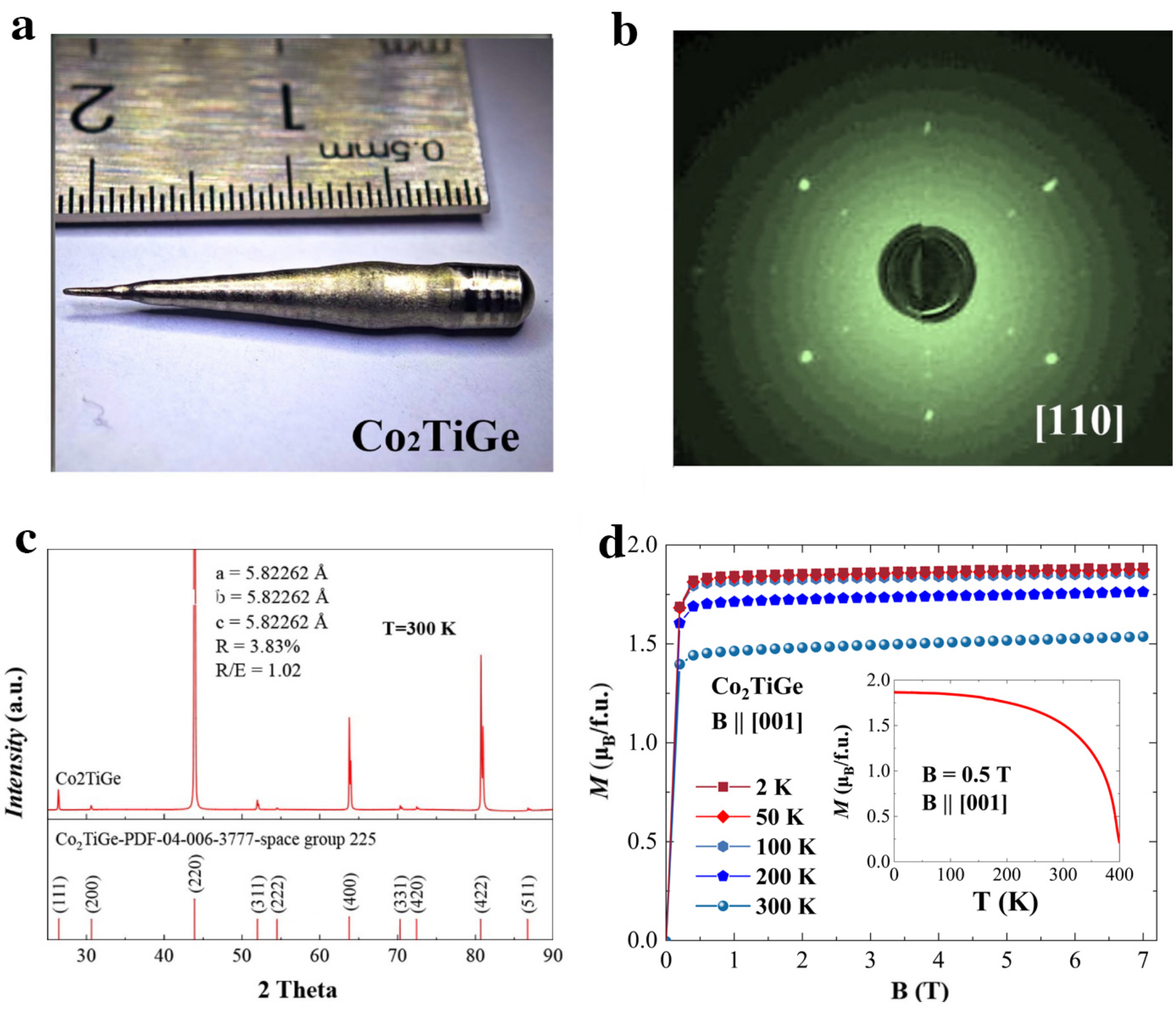

**Extended Data Fig. 2|Crystal growth, structural verification, and magnetic characterizations of $Co_2TiGe$ single crystals. a,** Photograph of a representative centimetre-scale $Co_2TiGe$ single-crystal boule synthesized via the ultra-high-temperature tetra-arc Czochralski method. **b,** Room-temperature X-ray Laue back-reflection pattern aligned along the $[110]$ crystallographic zone axis, exhibiting sharp, high-symmetry diffraction spots that verify pristine single crystallinity and the absence of multiple grains or twinning. **c,** Room-temperature powder X-ray diffraction (XRD) pattern obtained from pulverized single crystals, confirming high phase purity; all diffraction peaks index cleanly to the ordered $L2_1$ full-Heusler structure (space group $Fm\bar{3}m$ ) without parasitic secondary phases. **d,** Field-dependent isothermal magnetization curves $M(B)$ recorded at selected temperatures with the magnetic field aligned along the easy axis ($B \parallel [001]$), demonstrating typical soft-ferromagnetic behavior with high saturation magnetization. Inset: temperature-dependent magnetization $M(T)$ measured under an applied field of $\mu_0 H = 0.5$ T along the

[001] orientation, revealing a robust ferromagnetic phase transition with a Curie temperature of $T_C \approx 400$ K.

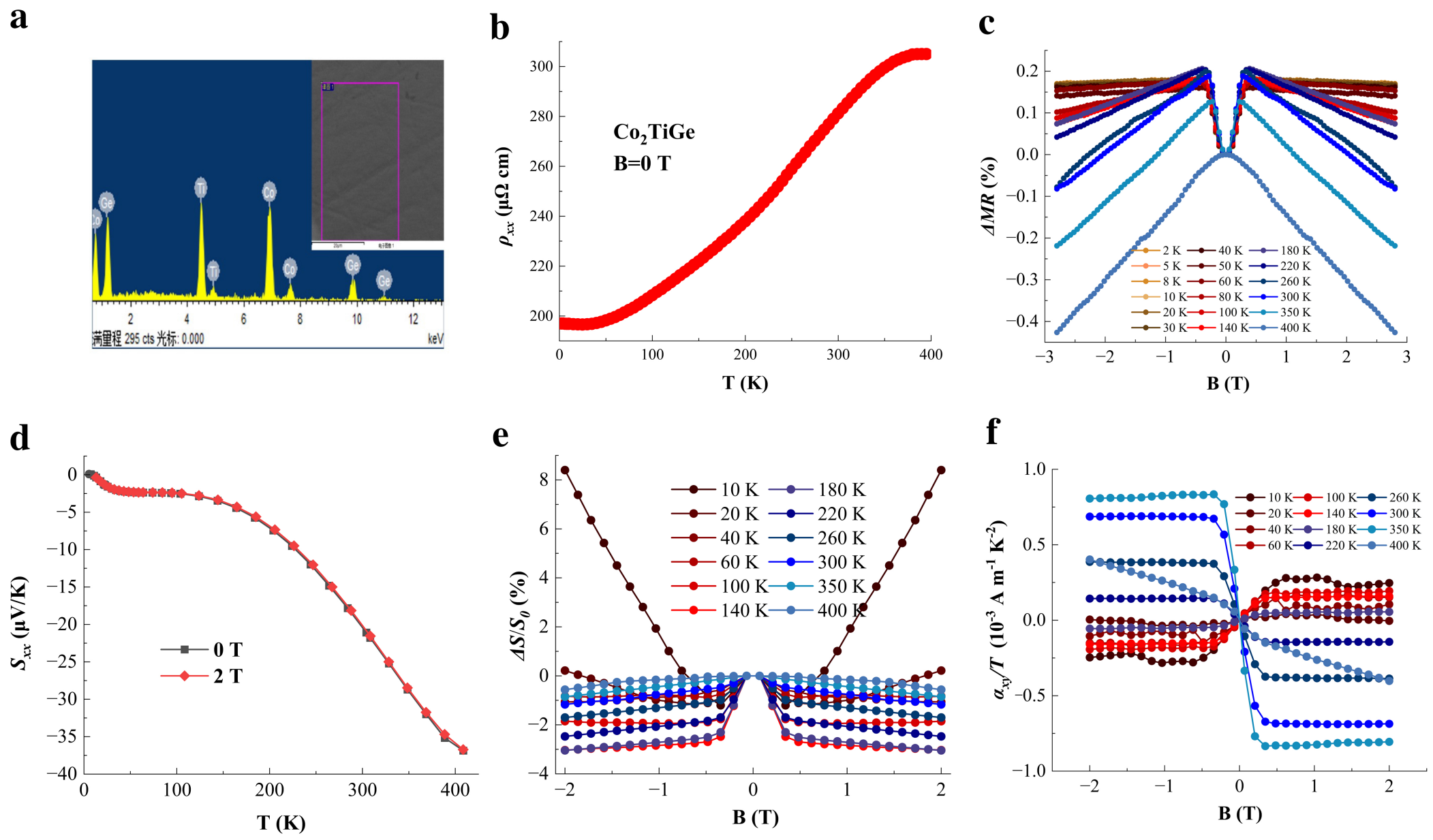

**Extended Data Fig. 3 |Compositional quantification and longitudinal magnetoelectric and thermoelectric transport characterizations of $Co_2TiGe$ single crystals. a**, Energy-dispersive X-ray spectroscopy (EDS) analysis collected via scanning electron microscopy (SEM) on a mechanically polished $Co_2TiGe$ single crystal. Quantitative statistical evaluation averaged over 20 independent spots yields an atomic ratio of Co:Ti:Ge=1.88(3) : 1.07(2) : 1.05(2), revealing a slight cobalt sub-stoichiometry that shifts of the actual Fermi level to $\boldsymbol{E_F}$=0.4473 eV. **b**, Temperature dependence of the zero-field longitudinal resistivity $\rho_{xx}(T)$, displaying characteristic metallic conduction across the entire temperature window down to 2 K. **c**, Transverse magnetoresistance $\Delta\rho_{xx}/\rho_0 = [\rho_{xx}(B) - \rho_{xx}(0)]/\rho_{xx}(0)$ under the orthogonal configuration **B** ⊥ **I** recorded at representative temperatures. **d**, Temperature dependence of the longitudinal Seebeck thermopower $S_{xx}(T)$ measured at $\mu_0 H =$ 0 Tand 2 T, demonstrating the weak sensitivity of the longitudinal channel to external magnetic fields. **e**, Magnetic field dependence of the magneto-Seebeck ratio $\Delta S_{xx}/S_0 = [S_{xx}(B) - S_{xx}(0)]/S_{xx}(0)$ measured across selected temperatures. **f**, Temperature evolution of the normalized anomalous Nernst conductivity $\alpha_{xy}/T$under a saturating field of $\mu_0 H = 2$ T, providing experimental benchmarks for quantitative

comparison with theoretical calculations.

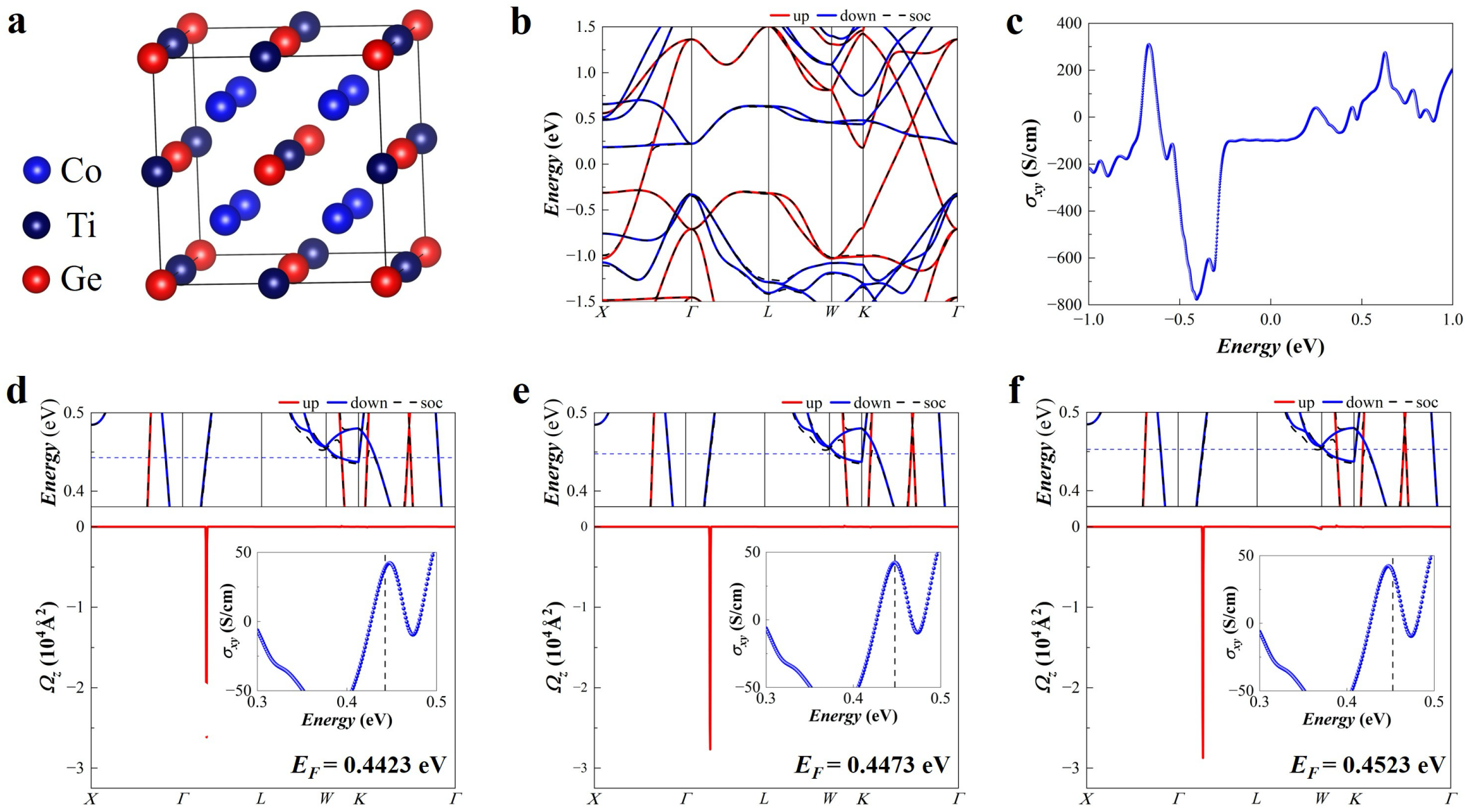

**Extended Data Fig. 4 | First-principles electronic band structure, AHC spectrum, and energy-dependent Berry curvature evolution.** **a**, Schematic crystal structure of full-Heusler cubic space group $Fm\bar{3}m$ (No. 225). Cyan, navy, and red spheres represent Co, Ti, and Ge atoms, respectively. **b**, Calculated electronic band dispersion along high-symmetry paths of the Brillouin zone. Red and blue curves designate the majority (spin-up) and minority (spin-down) bands in the absence of spin–orbit coupling (SOC), respectively; solid black curves delineate the fully relativistic band structure with SOC included. **c**, Energy-dependent intrinsic AHC spectrum $\sigma_{xy}(\varepsilon)$ calculated at zero temperature within a chemical potential window of $E_\mathrm{F}^0 \pm 1.0$ eV. Prominent topological resonance peaks are positioned at approximately $+0.2480$ and $+0.4473$, relative to the nominal Fermi level ($E_\mathrm{F}^0 = 0$ eV). **d–f**, Momentum-resolved Berry curvature distributions $\boldsymbol{\Omega}_z(\boldsymbol{k})$mapped along high-symmetry paths at three finely stepped chemical potentials across the targeted resonance peak: $\varepsilon = 0.4423$ eV ($E_0 -$ 5 meV; **d**), 0.4473 eV($E_0$; **e**), and 0.4523 eV($E_0 + 5$ meV; **f**). A colossal negative peak exceeding $-2.8 \times 10^4$ Å$^2$emerges sharply along the $\boldsymbol{\Gamma}$–$\boldsymbol{L}$ path, originating from an avoided band crossing driven by SOC. Insets show the corresponding chemical potential positions superimposed directly onto the steep slopes of the $\sigma_{xy}(\varepsilon)$ resonance peak, illustrating the microscopic origin of the sign reversal in the anomalous

Nernst conductivity.

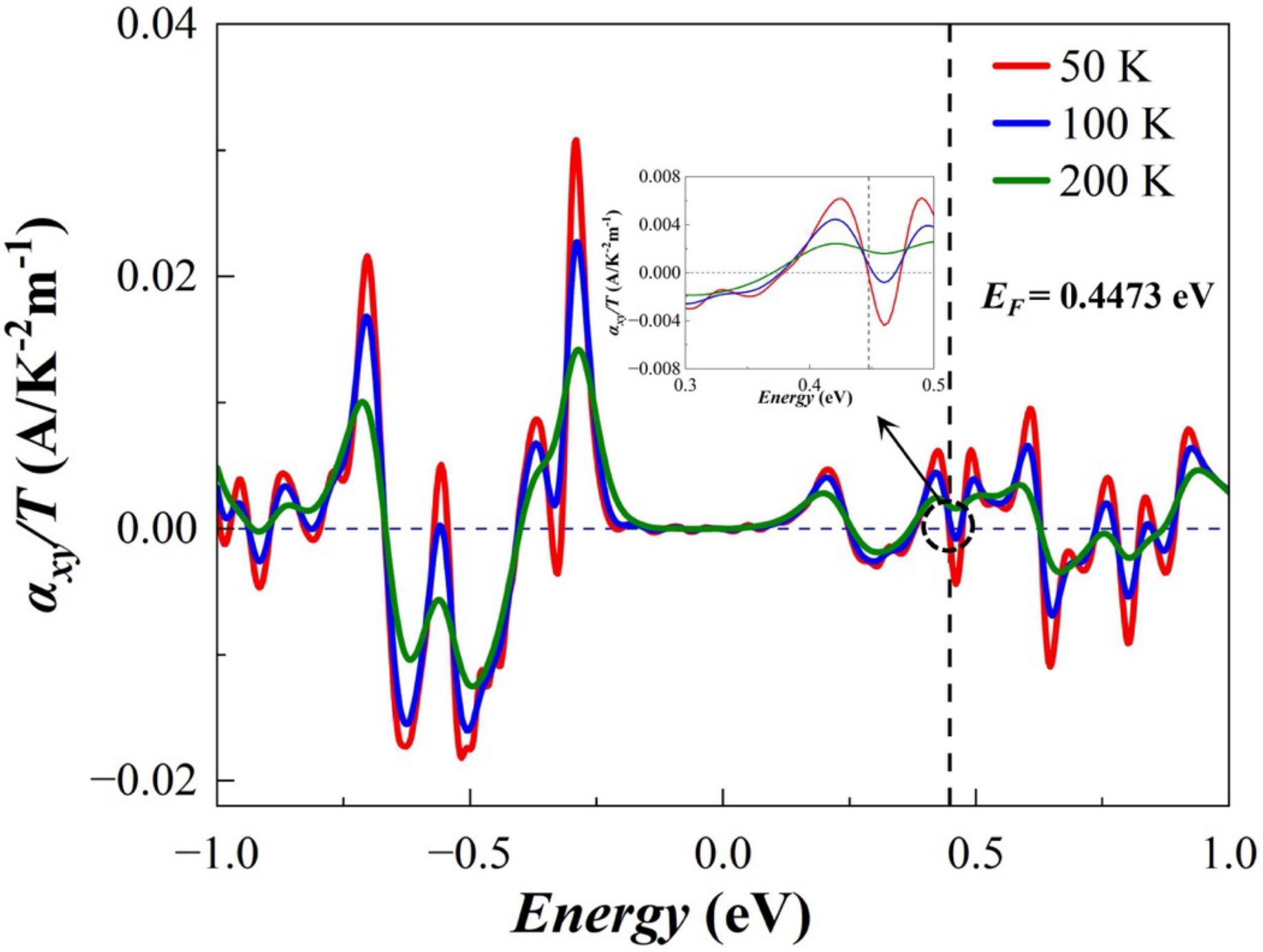

**Extended Data Fig. 5 |Energy dependence of the anomalous Nernst response.** Temperature-dependent anomalous Nernst response divided by temperature, $\alpha_{xy}/T$, as a function of energy at 50, 100, 200 K. The vertical dashed line marks $E_{\mathrm{F}} = 0.4473$ eV. The inset provides an enlarged view of $\alpha_{xy}/T$.

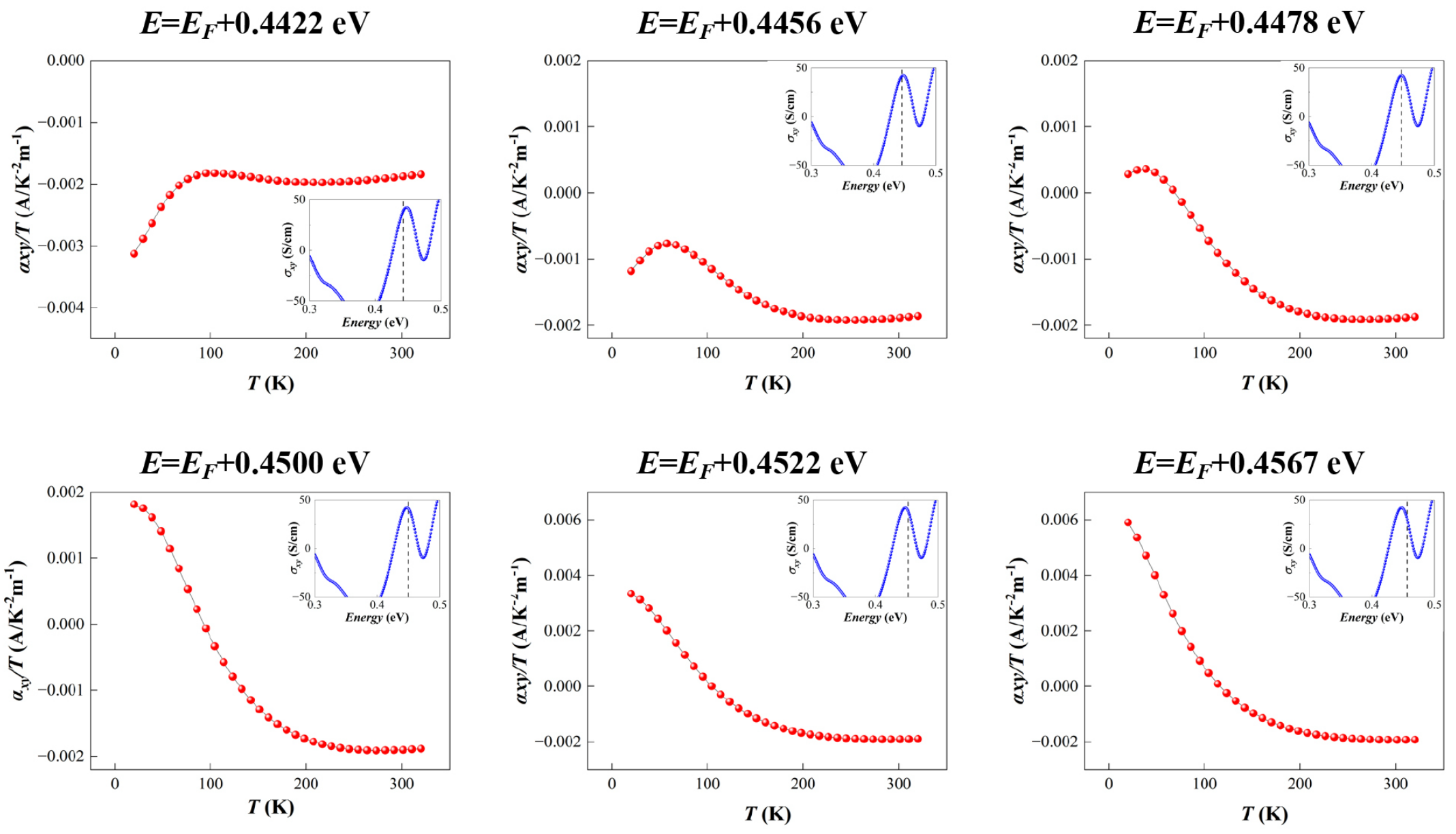

**Extended Data Fig. 6 |Temperature dependence of the anomalous Nernst response in $Co_2$TiGe. a–f**, Temperature dependence of $\alpha_{xy}/T$ at $E - E_{\mathrm{F}} = 0.4422$, 0.4456, 0.4478, 0.4500, 0.4522 and 0.4567 eV, respectively. Insets show the energy-dependent AHC $\sigma_{xy}(\varepsilon)$ with vertical dashed lines indicating the corresponding energies.

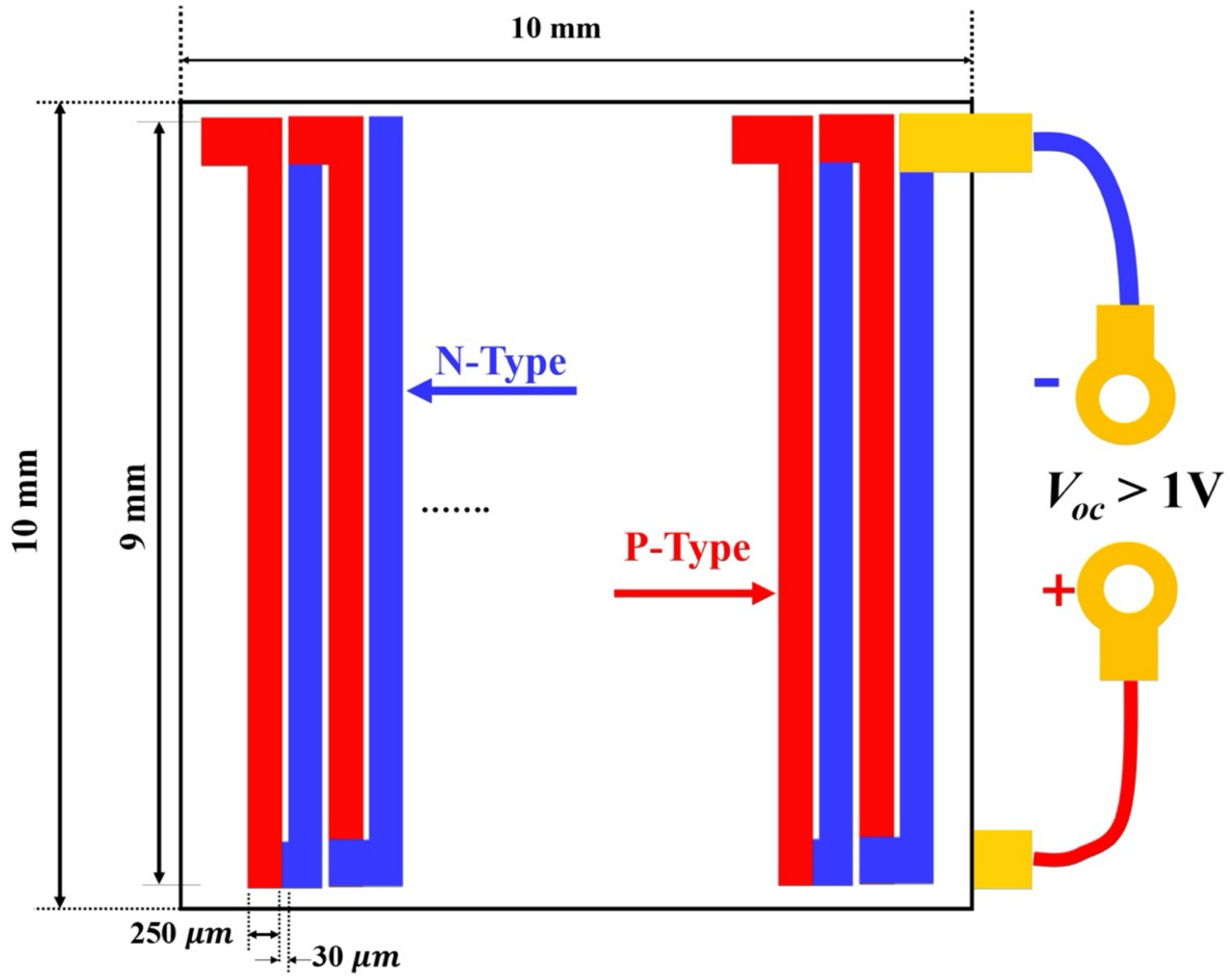

**Extended Data Fig. 7 | Structural layout, thermoelectric modeling, and device performance metrics of the dual-sided BANE thermopile.**

**Material Properties:** The thermoelectric platform utilizes the full Heusler compound $Co_2TiGe$ Experimentally, single-phase thin films exhibit a room-temperature anomalous Nernst coefficient of $S_{\text{ANE}}^{N-type} = -0.7\ \mu\text{V/K}$ along with an exceptionally low longitudinal resistivity of $\rho = 0.2\ \text{m}\Omega \cdot \text{cm}\ \ (2.0 \times 10^{-6}\ \Omega \cdot \text{m})$. To circumvent the fundamental trade-off imposed by non-active parasitic interconnects, we hypothesize the realization of a counter-doped, positive-polarity counterpart via isoelectronic or chemical-potential modulation (e.g., slight off-stoichiometric tuning of the transition metal/group-IV ratio), yielding $S_{\text{ANE}}^{P-type} = +0.7\ \mu\text{V/K}$ while preserving identical electronic transport metrics ($\rho = 2.0 \times 10^{-6}\ \Omega \cdot \text{m}$). The simultaneous access to *p- and n*-type ANE branches allows direct atomic-overlap planar junctions, eliminating internal potential cancellation under uniform out-of-plane temperature gradients.

**Planar and 3D Volumetric Geometric Layout:** The design targets a monolithic $1\ \text{cm} \times 1\ \text{cm}$ flexible polyimide (PI) substrate (nominal thickness $d_{\text{sub}} \approx 25\ \mu\text{m}$):

**Microstrip Dimensions:** Active microstrip width: $b = 250\ \mu\text{m} = 2.5 \times 10^{-4}\ \text{m}$; Thin-film thickness: $c = 1.0\ \mu\text{m} = 1.0 \times 10^{-6}\ \text{m}$; Cross-sectional area: $A_{\text{cs}} = b \cdot c = 2.5 \times 10^{-10}\ \text{m}^2$; Inter-strip isolation spacing (gap): $g = 30\ \mu\text{m} = 3.0 \times 10^{-5}\ \text{m}$; Array lateral pitch: $p = b + g = 280\ \mu\text{m}$.

**Array Integration per Surface:** Within a macroscopic active width $W_{\text{act}} = 8.93\ \text{mm}$, each substrate face accommodates $N = 32$ parallel strips (forming 16 thermocouple pairs of alternating positive and negative polarity): $W_{\text{act}} = N \cdot b + (N-1) \cdot g = 32 \times 0.25\ \text{mm} + 31 \times 0.03\ \text{mm} = 8.93\ \text{mm}$. To maintain a $1{:}1$ symmetric square active generation zone, the strip conduction length is established at $L = 8.94\ \text{mm}$ (yielding an active zone of $8.93\ \text{mm} \times 8.94\ \text{mm} \approx 0.80\ \text{cm}^2$ ). The peripheral zone provides a $\approx 0.5$ mmclearance boundary across all edges for fixture clamping and gold lead bonding.

**Dual-Sided 3D Series Topology:** Both top and bottom faces are patterned with identical 32-strip serpentines. An edge-through, gold-metallized micro-via (diameter $\approx 50\ \mu\text{m}$) links the negative terminal of the top circuit to the positive terminal of the bottom circuit in series. Single-surface conduction length: $L_{\text{side}} = N \cdot L = 32 \times 8.94\ \text{mm} = 286.08\ \text{mm} \approx 28.61\ \text{cm}$; Total effective active length: $L_{\text{total}} = 2 \times L_{\text{side}} = 572.16\ \text{mm} \approx 57.22\ \text{cm}\ (0.572\ \text{m})$.

**Thermoelectric and Electrical Derivation:** The operating conditions assume an applied vertical temperature difference $\Delta T = 10\ \text{K}$ across the film thickness $c$ ($\nabla T_z = \Delta T / c = 10^7\ \text{K/m}$) and saturated in-plane magnetization ($M \parallel y$).

**Open-Circuit Voltage ($V_{\text{oc}}$):** The generalized anomalous Nernst field is expressed as $E_{\text{ANE}} = -S_{\text{ANE}}(\nabla T \times m)$. The induced transverse electrical potential along a single microstrip segment of length $L$is:

$$V_{\text{strip}} = |S_{\text{ANE}}| \cdot \nabla T_z \cdot L = |S_{\text{ANE}}| \cdot \frac{\Delta T}{c} \cdot L$$

Substituting the structural and transport values:

$$V_{\text{strip}} = (0.7 \times 10^{-6}\ \text{V/K}) \times \left(\frac{10\ \text{K}}{1.0 \times 10^{-6}\ \text{m}}\right) \times (8.94 \times 10^{-3}\ \text{m})$$

$$= 6.258 \times 10^{-2}\ \approx 62.58\ \text{mV}$$

For $N = 32$ alternated strips interconnected in series on a single side:

$$V_{\text{single}} = \sum_{i=1}^{32} V_{\text{strip},i} = 32 \times 62.58\ \text{mV} \approx 2.00\ \text{V}$$

Because the out-of-plane thermal flux and in-plane magnetization uniformly penetrate the stack, the bottom array experiences identical transport driving forces. Under series bridging via the conductive via:

$$V_{\text{oc}} = V_{\text{top}} + V_{\text{bottom}} = 2 \times V_{\text{single}} = 2 \times 2.00\ \text{V} = 4.00\ \text{V}$$

**Device Internal Resistance ($R_{\text{int}}$):** Neglecting the negligible contact resistance of the wide overlap pads ( $50\,\mu\text{m} \times 50\,\mu\text{m}$ ) and the through-hole via, the longitudinal resistance per surface is governed by the thin-film channel geometry:

$$R_{\text{single}} = \rho \frac{L_{\text{side}}}{A_{\text{cs}}} = (2.0 \times 10^{-6}\ \Omega \cdot \text{m}) \times \frac{0.2861\ \text{m}}{2.5 \times 10^{-10}\ \text{m}^2} \approx 2289\,\Omega$$

For the combined two-sided series circuit:

$$R_{\text{int}} = 2 \times R_{\text{single}} = 2 \times 2288.64\ \Omega \approx 4578\,\Omega\ (\approx 4.58\ \text{k}\Omega)$$

**Maximum Power Delivery ($P_{\text{max}}$) and Areal Power Densities:**

Under conjugate matched load conditions ($R_{\text{load}} = R_{\text{int}}$), the maximum harvestable electrical power is calculated as:

$$P_{\text{max}} = \frac{V_{\text{oc}}^2}{4R_{\text{int}}} = \frac{(4.00\ \text{V})^2}{4 \times 4578\ \Omega} = \frac{16.0\ \text{V}^2}{18312\ \Omega} \approx 8.74 \times 10^{-4}\ \text{W} = 0.874\ \text{mW}$$

Normalizing this performance to spatial metrics:

**Active-area areal power density ($\omega_{\text{act}}$):**

$$\omega_{\text{act}} = \frac{P_{\text{max}}}{A_{\text{active}}} = \frac{0.874\ \text{mW}}{8.93\ \text{mm} \times 8.94\ \text{mm}} = \frac{0.874\ \text{mW}}{0.798\ \text{cm}^2} \approx 1.10\ \text{mW/cm}^2$$

**Total substrate areal power density ($\omega_{\text{sub}}$):**

$$\omega_{\text{sub}} = \frac{P_{\text{max}}}{A_{\text{substrate}}} = \frac{0.874\ \text{mW}}{1.0\ \text{cm} \times 1.0\ \text{cm}} \approx 0.87\ \text{mW/cm}^2$$

**Extended Data Table I |Summary of Structural and Electrical Specifications**

| Parameter | Symbol | Value | Unit |
|---|---|---|---|
| **Material Properties ($Co_2TiGe$)** | | | |
| Positive anomalous Nernst coefficient (doped) | $S_{ANE}^{P\text{-}Type}$ | +0.7 | $\mu$V/K |
| Negative anomalous Nernst coefficient (int-rinsic) | $S_{ANE}^{N\text{-}Type}$ | −0.7 | $\mu$V/K |
| Room-temperature resistivity | $\rho_{xx}$ | 0.2 | mΩ · cm |
| **Geometric Parameters** | | | |
| Substrate dimensions | $W_{sub} \times L_{sub}$ | 10.0 × 10.0 | $mm^2$ |
| Substrate material | — | Polyimide (PI) | — |
| Thin-film thickness | $c$ | 1.0 | $\mu$m |
| Microstrip width | $b$ | 250 | $\mu$m |
| Inter-strip isolation gap | $g$ | 30 | $\mu$m |
| Microstrip length | $L$ | 8.94 | mm |
| Microstrip cross-sectional area | $A_{cs}$ | $2.5\times10^{-10}$ | $m^2$ |
| Number of strips per surface | $N$ | 32 (16 pairs) | — |
| Single-surface active area | $W_{act} \times L_{act}$ | 8.93 × 8.94 | $mm^2$ |
| Single-surface cumulative active length | $L_{side}$ | 28.61 | cm |
| Total dual-surface effective length | $L_{total}$ | 57.22 | cm |
| Direct overlap junction area | $A_{pad}$ | 50 × 50 | $\mu m^2$ |
| **Operating Conditions** | | | |
| Out-of-plane temperature difference | $\Delta T$ | 10 | K |
| Out-of-plane temperature gradient across film | $\nabla T_z$ | $1.0\times10^{-7}$ | K/m |
| **Electrical Outputs & Performance** | | | |
| Single-strip output voltage | $V_{strip}$ | 62.58 | mV |
| Single-surface open-circuit voltage | $V_{single}$ | 2.00 | V |
| **Total open-circuit voltage** | $\mathbf{V_{oc}}$ | 4.00 | V |
| Single-surface internal resistance | $R_{single}$ | 2289 | Ω |
| Total internal resistance | $R_{int}$ | 4578 | Ω |
| **Maximum output power** | $\mathbf{P_{max}}$ | 0.874 | mW |
| **Active area power density** | $\boldsymbol{\omega_{act}}$ | **1.10** | **$mW/cm^2$** |
| Substrate area power density | $\omega_{sub}$ | 0.87 | $mW/cm^2$ |

# Supplementary for: Flat-band driven bipolar anomalous Nernst effect towards volt-scale transverse thermoelectric conversion

Haofu Wen[1]*, Xiaodong Guo[2]*, Xiaokang Xu[1]*, Peng Di[1]*, Xuebiao Ji[1], Ruoqi Wang[1], Shijun Yuan[1], Kunquan Hong[1], Qionghua Zhou[1]✉, Qian Chen[1]✉, Taishi Chen[1]✉, Jinlan Wang[1]✉

## S1. Candidate material distributions and missing magnetic labels

To visualize the distribution of materials in the Materials Project (MP) database, a 481-dimensional representation was constructed, comprising 349 CHGNet-derived features and 132 Magpie descriptors. The descriptors were standardized, reduced to 50 principal components and projected into two dimensions using t-distributed stochastic neighbour embedding (t-SNE). Matching the ICSD identifiers of Materials Project (MP) entries with those in the Flatband database yielded 32,006 materials common to both databases. These materials were broadly distributed across the descriptor map (Supplementary Fig. 1a).

Of the 154,377 materials in the MP dataset, 25,981 (16.83%) had an unknown magnetic label. These materials were distributed across much of the descriptor map (Supplementary Fig. 1b), indicating that missing magnetic labels were a widespread issue in the database. Among the 3,836 topological flat-band metal candidates, 1,653 had existing magnetic labels, whereas 2,183 (56.91%) had unknown labels (Supplementary Fig. 1c).

## S2. Classification Model for magnetic and nonmagnetic

To provide magnetic labels for screening, an XGBoost classifier was developed that distinguishes magnetic (M, including Ferromagnetic, Ferrimagnetic, and Antiferromagnetic from nonmagnetic (NM) materials. M and NM were assigned to positive and negative classes, respectively. The labelled dataset from the Materials Project (MP) database contained 128,396 materials, comprising 65,838 M and 62,558 NM materials. Another 25,909 materials labelled unknown were reserved for prediction and excluded from training and evaluation.

Each material was represented by 481 descriptors: 349 CHGNet-derived structural and magnetic-moment descriptors and 132 Magpie compositional descriptors. The labelled dataset was randomly split by MP ID into training, validation and test sets of 89,876, 12,840 and 25,680 materials, respectively. Model selection used the training, test, and validation sets. The classification threshold was fixed at 0.5 before test evaluation. Hyperparameters are listed in Supplementary Table S2. The evaluation model was fitted only to the training set; a separate deployment model was refitted using all 128,396 labelled materials.

**S3. Model performance**

After 2,400 boosting rounds, training and validation log-loss values of 0.1037 and 0.1397, respectively, were obtained, with the validation loss levelling off during the later rounds (Supplementary Fig. 2a). A test-set ROC-AUC of 0.9877 (Supplementary Fig. 2b) and an average precision of 0.9884 were achieved. At the classification threshold of 0.5, a validation F1 score of 0.9469 and an accuracy of 94.60% were obtained. The corresponding test-set accuracy was 94.81%, with M-class precision, recall and F1 score of 0.9562, 0.9420 and 0.9490.

To assess performance within the screened topological and metallic candidate pool, the 306 labelled materials common to the original test set and the pool of 3,836 candidates were evaluated. A ROC-AUC of 0.9681 was obtained for this subset. At the threshold of 0.5, 277 of the 306 materials were correctly classified (Supplementary Fig. 2c). Across the full test set, predicted M-class probabilities were concentrated near zero for NM materials and near one for M materials, with some overlap between the distributions (Supplementary Fig. 2d).

**S4. The Scaling plot between Intrinsic AHC and Longitudinal Conductivity**

In this work, the intrinsic anomalous Hall conductivity (AHC) at finite temperatures is evaluated using the Kubo formula within the linear response framework: $\sigma_{xy}^{int.} =$

$-\frac{e^2}{\hbar}\sum_n \int \frac{d^3\boldsymbol{k}}{(2\pi)^3} f(\varepsilon_n(\boldsymbol{k})\Omega_{n,z}(\boldsymbol{k}))$, here $f(\varepsilon_n(\boldsymbol{k})) = [1 + \exp((\varepsilon_n(\boldsymbol{k}) - \mu)/k_\mathrm{B}T)]^{-1}$is the Fermi-Dirac distribution function at temperature $T$ with chemical potential $\mu$, and $\Omega_{n,z}(\boldsymbol{k})$is the $z$-component of the Berry curvature for the $n$-th band. The Berry curvature is microscopically defined as: $\Omega_{n,z}(\boldsymbol{k}) = -2\sum_{n'\neq n} \frac{Im[<u_{n,k}|\hat{u}_x|u_{n',k}><u_{n',k}|\hat{u}_y|u_{n,k}>]}{(\varepsilon_{n'}-\varepsilon_n)^2}$, where $|u_{n,\boldsymbol{k}}\rangle$ denotes the periodic part of the Bloch wave function, and $\hat{u}_{x,y} = \frac{1}{\hbar}\frac{\partial H(\boldsymbol{k})}{\partial k_{x,y}}$ are the velocity operators. The magnitude of the Berry curvature is predominantly governed by the energy difference $(\varepsilon_{n'}(\boldsymbol{k}) - \varepsilon_n(\boldsymbol{k}))$. In the vicinity of an accidental flat-band crossing, this denominator approaches zero ($\varepsilon_{n'} \approx \varepsilon_n$) over an extended momentum volume due to the weak dispersion of the bands. This dramatically amplifies $\Omega_{n,z}(\boldsymbol{k})$, thereby generating colossal, hypersensitive Berry curvature hotspots and driving a substantial intrinsic anomalous transport signal.

**S5. Methodology for Berry curvature hot spot searching in $Co_2TiGe$**

To identity the momentum-space origin of the extremum in AHC of $Co_2TiGe$, we performed a systematic search for Berry-curvature hot spots throughout the three-dimensional Brillouin zone (BZ) using Wannier-interpolated Hamiltonian. Rather than restricting the analysis to selected high-symmetry planes or high-symmetry paths, the entire BZ was sampled on 201×201×201 $\boldsymbol{k}$-point mesh.

At each k point, the Berry curvature of each individual band was evaluated from the Wannier Hamiltonian. For a given chemical potential $\mu$, the occupied-state Berry curvature was then constructed as

$$\Omega^{occ}{}_z(\boldsymbol{k};\mu) = \sum_n \Theta(\varepsilon_{n\boldsymbol{k}} - \mu)_z\Omega_{n\boldsymbol{k}}(\boldsymbol{k}),$$

where $\varepsilon_{nk}$ and $\Omega_{n\boldsymbol{k}}(\boldsymbol{k})$ denote the energy and the $z$ component of Berry curvature of band n, respectively. $\Theta$ is the zero-temperature occupation function. Since the intrinsic AHC is governed by the Brillouin-zone integral of the occupied-state Berry curvature, $\Omega^{occ}{}_z(\boldsymbol{k};\mu)$ was used as the quantity for identifying the dominant momentum-space contributions. At each chemical potential, the resulting distribution of $\Omega^{occ}{}_z(\boldsymbol{k};\mu)$ was searched for large-magnitude extrema, which were taken as candidate Berry-curvature hot spots. In particular, the search was carried out at $\mu$-$E_F$=0.4473 eV, where the calculated AHC exhibits a pronounced maximum.

To clarify the electronic origin of the dominant Berry-curvature hot spots, we further examine the band dispersions along off-symmetry momentum paths passing through the corresponding hot-spot. In total, eight pairs of symmetry-related hot spots are identified. Since these hot spots are symmetry equivalent, we select one representative pair for detailed analysis.

The two hot spots are located at $H_1$ = (0.269, 0.535, -0.110) and $H_2$ = (-0.269, -0.535, 0.110) = $-H_1$. The band structures are calculated along the $P_1$ - $H_1$ - $P_2$ and $P_3$ – $H_2$ – $P_4$ paths, respectively. Here, $P_1$ = (0.269, 0.535, -0.5), $P_2$ = (0.269, 0.535, 0.000), $P_3$ = (-0.269, -0.535, 0.000) and $P_4$ = (-0.269, -0.535, 0.500). all momentum coordinates are expressed in the reciprocal-lattice basis (**g1**, **g2**, **g3**), defined by ai·gj=2πδij. For the real-space primitive vector $a_1$ = (0, a, a), $a_2$ = (a, 0, a), $a_3$ = (a, a,0), the corresponding reciprocal-lattice vectors are $g_1$=(π/a)(-1,1,1), $g_2$=(π/a)(1,-1,1), $g_3$=(π/a)(1,1,-1). These off-symmetry paths directly resolve the local band structure underlying the Berry-curvature hot spots, enabling their origin to be attributed to the associated band crossings.

To quantify the contribution of the identified Berry-curvature hot spots to the AHC, we integrated the occupied-state Berry curvature over local regions surrounding all 16 Weyl points, including $H_1$, $H_2$, and evaluated their contributions relative to the corresponding full-BZ integral:

$$\eta(\delta) = \frac{\int_{V(\delta)} \Omega^{occ}{}_z(k,\mu) d^3k}{\int_{BZ} \Omega^{occ}{}_z(k,\mu) d^3k}$$

where $V(\delta)$ the combined local regions surrounding hot spots H1 and H2, defined by $\Delta k_i < \delta$ along each reciprocal-lattice direction gi. For δ = 0.05, 0.06, and 0.07, the corresponding contributions amount to 65.8%, 84.6%, and 105.4% of the full-BZ Berry-curvature integral, respectively.

## S6. Scoring methodology

**Scoring methodology** The candidate materials were evaluated using six criteria. Scores for each criterion were normalized to the highest value among the ten materials and expressed on a 0–100 scale, with higher scores indicating more favourable characteristics.

**Curie temperature** The score was proportional to the Curie temperature, $T_c$[1-5]. The value of 1000 K for $Fe_3Ga$ was determined experimentally in this work, whereas the value of 1024 K for $Co_2FeIn$ was taken from theoretical calculations[1].

**Synthesis feasibility** Empirical scores of 80 were assigned to Heusler alloys, except for $Fe_3Ga$, which was assigned a score of 50. $Fe_3Sn$ was assigned a score of 30. These assignments incorporated our experimental synthesis experience, with representative reports of thin-film growth and synthesis routes[5-14].

**Band flatness** The raw score was defined as $S_{flat} = 100/(1+W/25\ \mathrm{meV})$where W is the bandwidth of the selected flat-band segment[15]. This definition assigns higher scores to narrower bands.

**Flat-band proximity to the Fermi level** The raw score was defined as $S_{prox}=100/(1+|E_{mid}-E_f|/100\ \mathrm{meV})$ where E mid is the energy midpoint of the same segment used to evaluate band flatness[15]. Higher scores correspond to flat bands closer to $E_f$.

**Topological screening** All ten materials satisfied both Flatband topological criteria: the best flat band belongs to a topological set, and all bands up to the best flat band belong to a topological set[15]. A score of 100 was therefore assigned to each material.

**Maximum absolute ANC** For the Noky dataset[16], the score was proportional to the largest absolute anomalous Nernst conductivity available at 300 K within $E_f$ ±0.5 eV, expressed in A $m^{-1}$ $K^{-1}$[17-21].

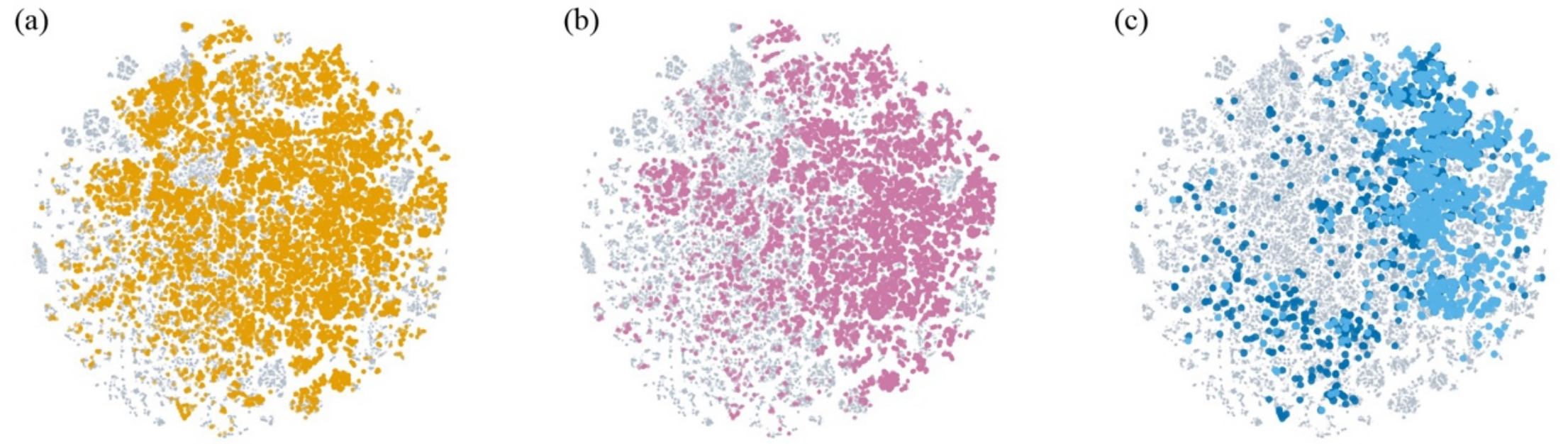

**Supp. Fig. 1 | Distribution of materials data and missing magnetic labels.**

All panels share the same two-dimensional t-SNE projection of the 481-dimensional representation comprising 349 CHGNet-derived features and 132 Magpie descriptors. Light-grey points represent the 154,305 MP materials with available descriptors. **a,** Materials shared by MP and the Flatband database through exact ICSD identifier matching (orange; 32,002 plotted materials). **b,** MP materials with unknown magnetic labels (pink; 25,909 plotted materials). **c,** The 3,836 topological flatband metal candidates, comprising 1,653 materials with existing magnetic labels (dark blue) and 2,183 with unknown labels (light blue).

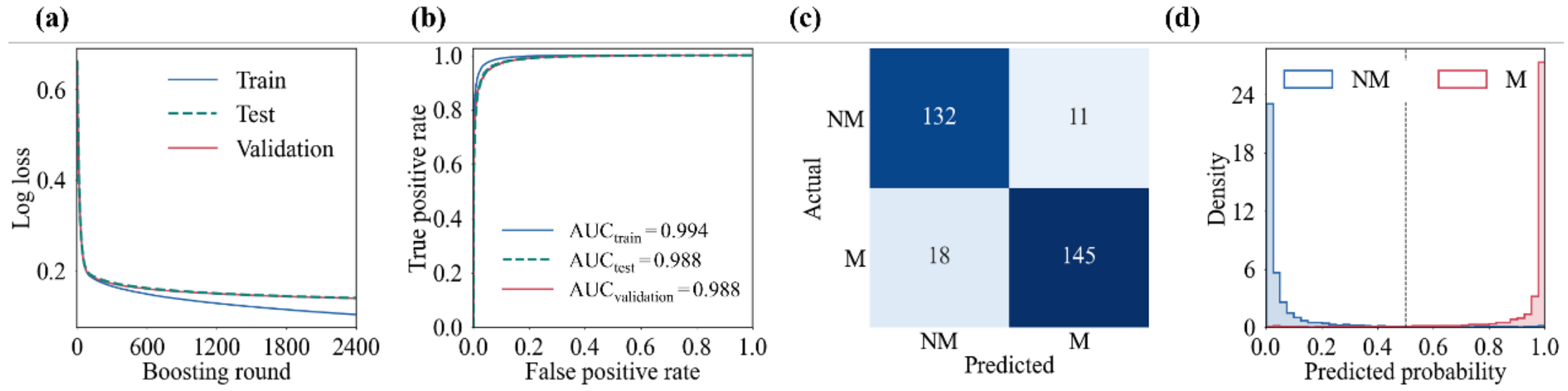

**Supp. Fig. 2 |Performance of the magnetic classification mode. a,** Binary log-loss for the training, test and validation sets over 2,400 boosting rounds. **b,** ROC curves for the training, test and validation sets, with AUC values of 0.994, 0.988 and 0.988, respectively. **c,** Confusion matrix for 306 labelled candidate-pool test materials at a threshold of 0.5, where M denotes magnetic materials (Ferromagnetic, Ferrimagnetic, and Antiferromagnetic) and NM denotes nonmagnetic materials. **d,** Predicted M-class probability distributions for NM and M materials in the test set; the dashed line marks the threshold of 0.5**.**

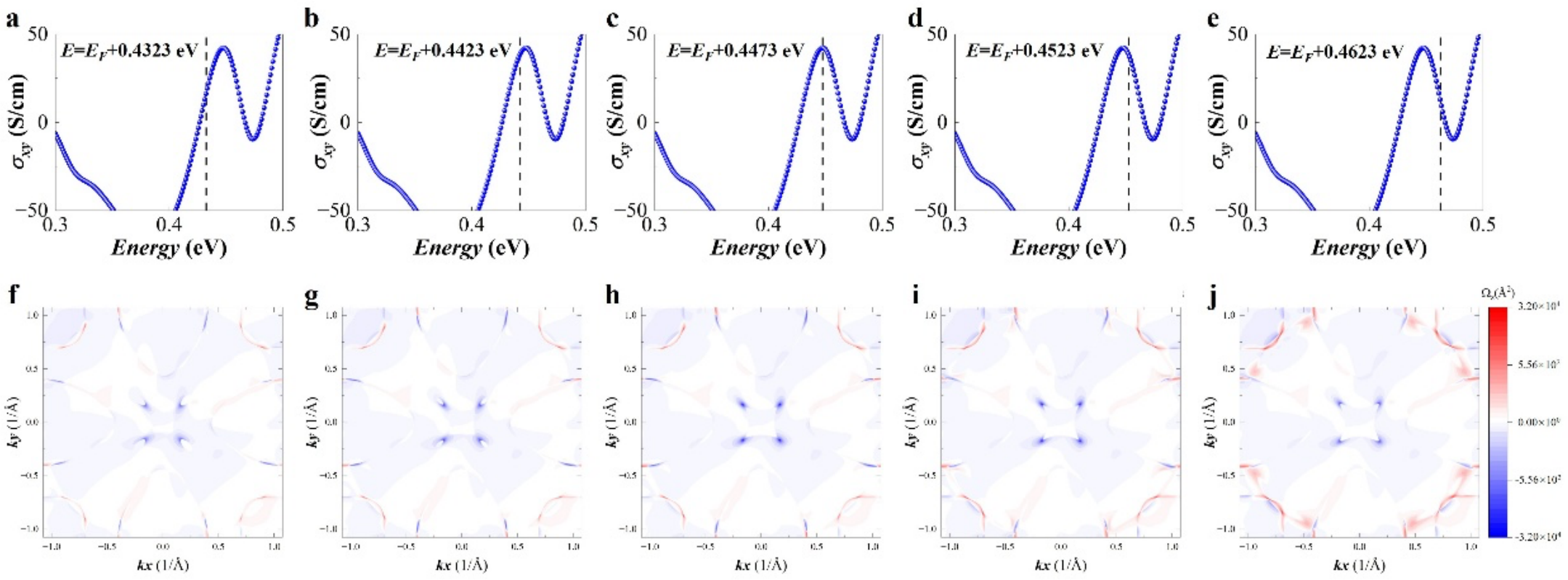

**Supp. Fig. 3 Energy-dependent Berry curvature distribution in kz = 0.155 (2π/a) plane. a-e** the corresponding energy-dependent anomalous Hall conductivity $\sigma_{xy}(\varepsilon)$, with dashed vertical lines indicating the selected energies of $E$-$E_F$ = 0.4323, 0.4423, 0.4473, 0.4523, and 0.4623 eV, respectively. **f-j** the lower panels show momentum-resolved Berry curvature $\Omega_z$ (***kx***, ***ky***) plane at selected energies.

## Table S1. Overall materials-screening workflow

| Step | Screening criterion | Retained materials (unique MP IDs) |
|---|---|---|
| 0 | Full Materials Project v2026.04.13 dataset | 154,377 |
| 1 | Exact intersection with experimental ICSD identifiers in the Materials Flatband Database | 32,006 |
| 2 | Flatband filtering with both topological-feature flags enabled | 5,023 |
| 3 | Materials Project is_metal = True | 3,836 |
| 4 | Original MP magnetic-ordering labels were prioritized; model predictions were applied only to entries labeled Unknown; only the M class was retained | 1,211 |
| 5 | Strict priority screening followed by manual curation | 249 |

## Table S2. Hyperparameters of XGBoost classifier

| Hyperparameter | Value |
|---|---|
| Boosting rounds | 2,400 |
| max_depth | 3 |
| learning_rate | 0.04 |
| subsample | 0.75 |
| colsample_bytree | 0.75 |
| min_child_weight | 20 |
| reg_lambda | 20 |
| reg_alpha | 1 |
| gamma | 0.2 |

## Table S3. Flatband parameters for strict priority screening

| **Category** | **Materials Flatband Database filter** | **Setting** |
|---|---|---|
| Materials features | Point group | All |
| Materials features | Space group | All |
| Materials features | Minimum number of distinct elements | 1 |
| Materials features | Maximum number of distinct elements | 10 |
| Materials features | Maximum number of electrons | 400 |
| Materials features | Required chemical elements | None; |
| Flat-band features | Require a flat band near $E_f$ | Enabled |
| Flat-band features | Energy window around $E_f$ | 0.5 eV |
| Flat-band features | Minimum segment length | 10 |
| Flat-band features | Maximum bandwidth for a flat segment | 0.025 eV |
| Flat-band features | Maximum number of bands without SOC in the energy window | 9 |
| Flat-band features | Minimum best-flatband score | 0.02 |
| DOS features | Require a DOS peak near $E_f$ | Enabled |
| DOS features | Maximum distance between $E_f$ and the nearest DOS peak | 0.25 eV |
| DOS features | Maximum Gaussian width of the DOS peak closest to $E_f$ | 0.75 eV |
| Topological features | Best flat band should belong to a topological set | Enabled |
| Topological features | All bands up to the best flat band should belong to a topological set | Enabled |
| Lattice features | Bipartite sublattice | Disabled |
| Lattice features | Pyrochlore sublattice | Disabled |
| Lattice features | Kagome sublattice | Disabled |
| Lattice features | Lieb sublattice | Disabled |
| Lattice features | Split sublattice | Disabled |
| Lattice features | Line-graph sublattice | Disabled |
| Tags | Manually checked as flat band | Disabled |
| Tags | Manually checked as flat atomic band | Disabled |
| ICSD options | Include theoretical ICSD entries | Disabled |
| ICSD options | Display all ICSD entries | Enabled |